\documentclass[twoside,11pt]{article}

\usepackage{natbib}

\usepackage{amsmath}
\usepackage{amsthm}
\usepackage{url}
\usepackage{graphicx}
\usepackage{graphics}
\usepackage{epsfig}
\usepackage{amsmath}
\usepackage{multirow}
\usepackage{subfigure} 
\usepackage{rotating}
\usepackage{amssymb}
\usepackage{wrapfig} 
\usepackage{enumerate}
\usepackage{times}
\usepackage{caption}
\usepackage{authblk}

\usepackage[ruled]{algorithm2e}

\begin{document}

\title{Toward a Robust Crowd-labeling  Framework using \\Expert Evaluation and Pairwise Comparison}

\author[1]{ Faiza Khan Khattak}
\author[2]{ Ansaf Salleb-Aouissi}
\affil[1,2]{Department of Computer Science. Columbia University. New York}



\maketitle

\begin{abstract}
{\em Crowd-labeling} emerged from the need to label large-scale and complex data, a tedious, expensive, and time-consuming task. One of the main challenges in the crowd-labeling task is to control for or determine in advance the proportion of low-quality/malicious labelers. If that proportion grows too high, there is often a {\em phase transition}  leading to a steep, non-linear drop in labeling accuracy as noted by \protect \cite{Karger}. To address these challenges, we propose a new framework called Expert Label Injected Crowd Estimation (ELICE)  and extend it to  different versions and variants  that delay phase transition leading to a better labeling accuracy. ELICE automatically combines and boosts bulk crowd labels supported by labels from experts for limited number of instances from the dataset. The expert-labels help to estimate the individual ability of crowd labelers and  difficulty of each instance, both of which are used to aggregate the labels.
 Empirical evaluation shows the superiority of ELICE as compared to other state-of-the-art methods. We also derive a lower bound on the number of expert-labeled instances needed to estimate the crowd ability and dataset difficulty as well as to get better quality labels.
\end{abstract}

\section{ Introduction}
\label{introduction}
\vspace*{0.2cm}

Crowd-labeling is the process of having a human crowd label a large dataset --  a tedious, time-consuming and expensive process if accomplished by experts alone. 
The idea of using crowd intelligence is not new \cite{Nelson} but nowadays it is accomplished in a more sophisticated way by publishing the tasks on the web. One example of a successful crowd-labeling system is the reCAPTCHA project for digitizing old books  \cite{recaptcha2008}. This project  leverages the power of human volunteers to transcribe approximately 500 million words at close to 100\% accuracy, words that were otherwise unrecognizable by Optical Character Recognition (OCR) software. Another example of a widely used crowd-labeling system is Amazon's Mechanical Turk (AMT), which engages thousands of workers registered to apply their brainpower to complex labeling tasks. These include for instance, annotating medical images that contain malignant cells and identifying videos suitable for a general audience.

\subsection{Challenges and Phase Transition}
\vspace*{0.2cm}
 In a crowd-labeling scene, an object is usually annotated by more than one labeler. The multiple labels obtained per object are then combined to produce one final label for quality assurance. 
Since the ground truth, instance difficulty and the labeler ability are generally unknown entities, the aggregation task becomes a ``chicken and egg'' problem to start with. While  significant progress has been made on the process of aggregating crowd-labeling results, e.g., \cite{Karger,Sheng2008,WhitehillNIPS2009}, it is well-known that the precision and accuracy of labeling can vary due to differing skill sets. The labelers can be \emph{good/experienced}, \emph{random/careless} or even \emph{malicious}. 

Malicious labelers can include both  intentionally or unintentionally malicious. Intentionally malicious labelers are those who identify the correct labels and change them strategically while unintentionally malicious labelers demonstrate the same labeling behavior due to some misunderstanding about the labeling task. Throughout the rest of this paper, we will refer to both kinds as malicious labelers.

\begin{figure}[h]
\begin{center}
\vspace{-4cm}
\includegraphics[scale=0.55]{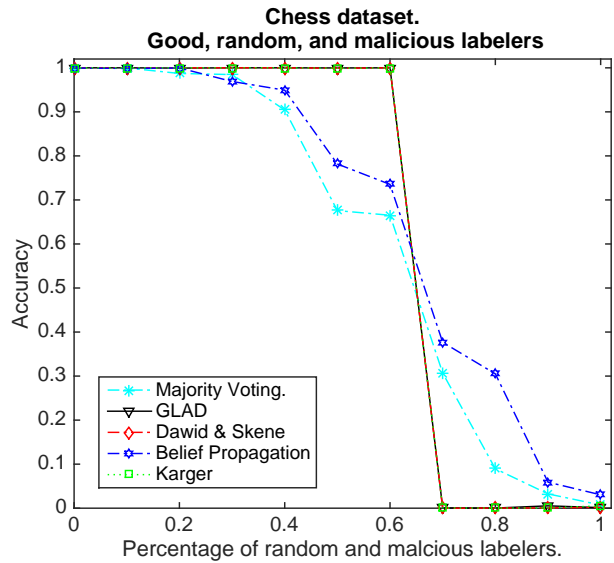}
\vspace{-4cm}
\caption{Phase transition in the performance of majority voting, GLAD \protect \cite{WhitehillNIPS2009}, Dawid and Skene's method \protect \cite{Dawid:Skene:79}, Belief Propagation \protect \cite{Qiang} and Karger's iterative method \protect \cite{Karger} on the UCI Chess dataset.}
 \end{center}
\label{phase}
 \end{figure}

One of the main challenges in crowd-labeling is that the proportion of low-quality/malicious labelers is unknown. High proportion of low quality (random and malicious) labelers can often result into a {\em phase transition} leading to a steep, non-linear drop in labeling accuracy as noted by  \cite{Karger}. 

We observed a similar phenomenon in the experiments we conducted on five UCI datasets \cite{AsuncionNewman:2007}. We used majority voting, GLAD (Generative model of Labels, Abilities, and Difficulties) by \protect \cite{WhitehillNIPS2009}, Dawid and Skene's method \protect \cite{Dawid:Skene:79}, Karger's iterative method \protect \cite{Karger} and Belief Propagation \cite{Qiang}. The labels for all these datasets were simulated. Figure 1 illustrates the phase transition for the UCI Chess dataset of 3,196 instances. We assume that a {\em good labeler} makes less 35\% mistakes, {\em random labeler} makes  between 35\% to 65\% mistakes, while a {\em bad labeler} makes more than 65\% mistakes. This highlights the larger challenge of producing an objective assessment to measure the quality of the crowd for a given task.

 Other than phase transition, many basic questions remain unresolved that make crowd-labeling a prevailing research topic, e.g., \cite{Dekel2009,MACE1,Qiang,Donmez_2008}. The unresolved questions include:

  \begin{enumerate}
  \item What are the best ways to evaluate labeler ability and instance difficulty? 
    \item It is common to use expert-labeled instances or ground truth to evaluate labelers and instances \cite{Edmonds,faiza_nips_2011,faiza_woc_2012,khattak13}. The question is, how many expert-labeled instances should be used in order to obtain an accurate evaluation?
     \item How can labelers and instances be evaluated if ground truth is not known with certitude?
    \item Is there any optimal way to combine multiple labels to get the best labeling accuracy?  
        \item Should the labels provided by malicious labelers be discarded and blocked? Or is there a way to use the ``information" provided by malicious labelers?
   \end{enumerate}

\subsection{Our Approach}
\vspace*{0.2cm}
In this paper, we investigate these questions. Building upon our preliminary work \cite{faiza_nips_2011,faiza_woc_2012,khattak13},
 we present a general framework called Expert Label Injected Crowd Estimation (ELICE). 
Earliest versions of ELICE were published in the workshop papers \cite{faiza_nips_2011,faiza_woc_2012}, which are presented as ELICE 1 in this paper. We also present ELICE 2  \cite{khattak13} with a {\em new and improved} aggregation method that genuinely takes advantage of the labels provided by malicious labelers. 

This paper summarizes our previous efforts to devise a robust crowd-labeling framework using both expert evaluation and pairwise comparison between crowd-labelers. 
Besides reporting the above-cited methods, we also propose an unpublished method ELICE 3 that handles the situation when the ground truth is not known with certainty. As a result, pairwise (or circular) comparison of labelers to labelers and instances to instances are incorporated to fill the gap. We also extend the experimental section by including in-depth experiments and comparisons with the latest state-of-the-art methods. 
 
 The goal of ELICE is to provide better accuracy for the labeling/annotation tasks for which predefined options of answers are available. We have assumed the scenario of labeling to be questions with multiple choices provided to the labelers. 
 
 All versions of ELICE rely on expert-labels for a small subset of randomly chosen  instances from the dataset. However, it can be noted that instead of random choice of the instances, experts can also help in identifying the representative instances of each class. These expert-labeled instances are used to evaluate labeler ability and data instance difficulty that help to improve the accuracy of the final labels.  For the first two versions of ELICE, we assume that expert-labels are ground truth labels. In the third version of ELICE however, we assume that expert-labels may not be ground truth either because the experts do not agree on the same label or because the instances are difficult and alternative methods to get ground truth are infeasible.


The first version of ELICE estimates the parameters, i.e., labeler expertise and data instance difficulty, using the accuracy of crowd labelers on expert-labeled instances \cite{faiza_nips_2011,faiza_woc_2012}. The multiple labels for each instance are combined using weighted majority voting. These weights are the scores of labeler reliability on any given instance, which are obtained by inputting the parameters in the logit function. In the second version of ELICE \cite{khattak13}, we introduce entropy as a way to estimate the uncertainty of labeling. This provides an advantage of differentiating between good, random and malicious labelers. The aggregation method for ELICE version 2 flips the label (for binary classification case) provided by the malicious labeler thus utilizing the information that is generally discarded by other labeling methods.

  Both versions of ELICE have a cluster-based variant in which rather than making a random choice of instances from the whole dataset, clusters of data are first formed using any clustering approach e.g., K-means. Then equal number of instances from each cluster are chosen randomly to get expert-labels. This is done to ensure equal representation of each class in the test-dataset.

Besides taking advantage of expert-labeled instances, the third version of ELICE, incorporates pairwise/circular comparison of labelers to labelers and instances to instances. The idea here is to improve the accuracy by using the crowd-labels, which unlike expert-labels, are available for the whole dataset and may provide a more comprehensive view of the labeler ability and instance difficulty. This is especially helpful for the case when the domain experts do not agree on one label and ground truth is not known for certain. Therefore, incorporating more information beyond expert-labels can provide better results.

We show empirically that our approaches are robust even in the presence of a large proportion of low-quality labelers in the crowd. This procedure also helps in stabilizing labeling process and delaying the phase transition to inaccurate labels. Furthermore, we derive a lower bound of the number of expert labels needed \cite{khattak13}. This lower bound is a function of the overall quality of the crowd and difficulty of the dataset.
 \medskip

 This paper is summarized and organized as follows:
\begin{itemize}
  \item We develop a new method called ELICE (three different versions with cluster-based variants), for aggregation of crowd labels based on a few expert-labeled instances. The aim of this method is to provide high accuracy of final label and delay in the phase transition.
  \item ELICE is designed for crowd-labeling tasks with predefined classes to choose from.
 \item A simple to implement, basic version of ELICE is introduced in Section \ref{elice_old}.
 \item Section \ref{elice_entropy} explains ELICE 2, which is capable of identifying the malicious labelers and adjusting their labels. This introduces a better way to aggregate the labels utilizing the information provided by malicious labelers.
 \item  Section \ref{elice_plus1} describes the methodology for ELICE 3, which uses the expert labels as well as pairwise comparison of labeler to labeler (and instance to instance) to gain maximum information from all sources when ground truth is not known for certain.
 \item In Section \ref{experiments},  we present the experimental evaluation of our methods on five UCI datasets and two real-life datasets. The labels for UCI datasets are simulated while the labels for real-life datasets are acquired by posting the tasks on Amazon Mechanical Turk (AMT).
 \item Section \ref{comparison} compares and discusses the different versions of ELICE.
 \item  In Section \ref{theorem2}, we present the theoretical framework to derive a lower bound on expert labels needed for ELICE.
 \item Section \ref{related_work} summarizes the related work in the crowd-labeling area.
 \item Section \ref{conclusion} concludes the paper.
 \end{itemize}

\section{ELICE 1 Framework}
\label{elice_old}
\vspace*{0.2cm}
The first version of ELICE \cite{faiza_nips_2011} is a simple yet effective and efficient method. Labeler ability is defined as the normalized value of the sum of all correct labels minus the number of incorrect labels provided for the expert-labeled instances. Similarly the difficulty of an expert-labeled instance is the normalized sum of the correct labels provided for that instance. The other instances for which  expert labels are not available are evaluated based on the estimated label. This estimated label is derived by majority voting weighted by each labeler ability. In the next step, labels are combined to form one final label. Another variant of the first version of ELICE, called ELICE with clustering is introduced in this section. The detailed methodology is described below.

\subsection{ELICE 1}
Consider a dataset of $N$ instances, which is to be labeled as positive (+1)  or negative (-1). A subset\footnote{This assumes that expert-labeled instances are representative of the available data.} of $n$ instances is labeled by an expert of the domain. There are $M$ crowd labelers who label all  $N$ instances. The label given to the  $i^{th}$ instance by the $j^{th}$ labeler is denoted by  $l_{ij}$. 

The expertise level of a labeler $j$  is denoted by $\alpha _j$, which can have a value between -1 and 1, where 1 is the score of a labeler who labels all instances correctly and -1 is the score of a labeler who labels everything incorrectly. This is because the expertise of a crowd labeler is penalized by subtracting 1 when he makes a mistake but it is incremented by 1 when he labels correctly. At the end, the sum is divided by $n$. Similarly, $\beta _i$ denotes the level of the difficulty of instance $i$, which is calculated by adding 1 when a crowd labeler labels that particular instance correctly. The sum is normalized by dividing by $M$. It can have value between 0 and 1, where 0 is for difficult instances and 1 is for the easy ones.  We calculate $\alpha_j $'s and  $\beta_i$'s  are as follows,\\
  \begin{equation} \label {alphbeta} \alpha_j=\frac{1}{n} \sum_{i=1}^n[{\bf 1}(L_i=l_{ij})-{\bf1}(L_i\neq l_{ij})]  \;\;\;\;\;\;\;\;\;\;\;\;\;\;\;\;\; \beta_i=\frac{1}{M} \sum_{j=1}^M[{\bf1}(L_i=l_{ij})]\end{equation}
   where        $  j=1,\ldots , M$  and   $ i=1,\ldots , n$.\\
   
   We infer the rest of  the $(N-n)$ number of $\beta$'s based on $\alpha$'s. As the true labels  for the rest of instances are not available, we try to find an approximation which we name as {\em expected label} (EL), 
    \begin{equation} \label{es_label} EL_i=sign(\frac{1}{M} \sum_{j=1}^M\alpha_j* l_{ij} ) \end{equation}

These expected labels are used to approximate $\beta$'s ,
     \begin{equation} \label{alphbeta2} \beta_i=\frac{1}{M} \sum_{j=1}^M[{\bf1}(EL_i=l_{ij})] \end{equation}
    The logistic function denoted by $\sigma$ is used to calculate the score associated with the correctness of a label, based on the level of expertise of the crowd labeler and the difficulty of the instance. This score gives us the approximation of the true labels called {\em inferred labels} (IL)  using the following formulas,
%
   \begin{equation} IL_i= sign(\frac{1}{M}\sum_{j=1}^M \sigma(\alpha_j \beta_i)* l_{ij})  \end{equation}

\subsection {ELICE 1 with Clustering}
We propose a variation of ELICE called ELICE with clustering. Instead of picking the instances randomly from the whole dataset $\cal{D} $ to acquire expert labels, clusters of instances  in $\cal{D}$ are first formed by applying k-means clustering using the features (if available); then equal numbers of instances are chosen from each cluster and given to the expert to label. This allows us to have expert labeled instances from different groups in the data, particularly when the dataset is highly skewed. Another possibility is to use any other method of clustering for instance K-means++ \cite{Arthur}.

\section{ELICE 2 Framework}
\label{elice_entropy}
\vspace*{0.2cm}

In the first version of ELICE, the random and malicious/adversarial labelers are treated in the same way i.e., their opinion is weighted less than the good labelers. But it is known from the crowd-labeling literature, e.g., \cite{RaykarY12} that malicious labelers can be informative in their own way and once they are identified, their labels can be adjusted to get the underlying possibly correct labels. 

 The random labelers on the other hand, are those who label without paying attention to instances. Therefore, their labels merely add noise to the labeling process. The malicious labelers are not random in their labels. They take time to identify the instance, try to infer the correct label and then flip it intentionally or unintentionally (assuming binary classification). Therefore, if we know the underlying intentions of a labeler in advance, we can obtain the correct label by decrypting the provided label. 
 
In the second version of ELICE, we have incorporated the idea of utilizing the labels provided by the malicious labelers. Just like the previous version of ELICE, the labeler ability and instance difficulty are evaluated but this time the evaluation involves the concept of entropy. Entropy measures the uncertainty of the information provided by the labelers (or uncertainty about the information obtained for the instances). A random labeler will have a high entropy while the good or malicious labeler will have a low entropy. 

This lets us differentiate between random vs. malicious or good labelers. Then the malicious and good labelers are separated. ELICE 2 assigns low weights to the labels of a random labeler and high weights to the labels of good labeler. Malicious labelers' annotations are also highly weighted but after adjusting the labels provided by them. This helps us in using the information that is discarded by many label aggregation methods. 
 Clustering method can also be used for this version of ELICE.


\subsection{ELICE 2}

Let $\cal{D}$ be a dataset of $N$ unlabeled instances. We assign $M$ crowd labelers to label the whole dataset; each instance $i$ will receive a label $L_{ij} \in \{  \pm 1 \}$ from labeler $j$, where $i \in \{1,\ldots,N\}$ and $j \in \{1,\ldots, M\}$. To evaluate the performance of the labelers, we get ``ground truth'' labels for a random sample $\cal{D'} (\subset \cal{D}) $ of cardinality $n<<N$. Instances of $\cal{D'}$ are labeled by one or more experts.


\subsubsection{\bf Labeler Expertise}

We use expert-labeled instances to evaluate the labelers by finding the probability of getting correct labels. This  estimation of labeler's  performance has a factor of uncertainty since it is based on a sample. Therefore, the entropy function can be a natural way to measure this uncertainty. 
Entropy is high when the probability is around 0.5 as we are least certain about such a labeler and it is low when the probability is close to 0 or 1. 
The formula for the entropy for a worker $j$  is given by:
 \begin{equation} E_j= - p_j log(p_j)-q_j log(q_j)\;\;\;\;\; \mbox{such that ,}\;\;\;\;\;\;p_j=\frac{\;n_j^+}{n}\;\;\;\;\;\;\;q_j=1- p_j \end{equation}
 \indent $n_j^+=| \mbox{correctly labeled instances from } \cal{D'} \mbox{ by labeler j}|$ \\

 Since we are more interested in the reliability of the assessment, we take $(1-E_j).$ In order to differentiate between good and bad labelers, we multiply by $(p_j-q_j)$. This assigns a  negative  value to the bad labeler and positive value to the good one.  We define the expertise of the labeler as
\begin{equation}\label{alpha}\alpha_j=(p_j-q_j)(1-E_j)\end{equation}\\
where $\alpha_j \in (-1,1)$. The multiplication by $(p_j-q_j)$ also allows for less variability in $\alpha_j$ when the number of correct and incorrect labels is close, assuming that it can be due to the choice of the instances in $\cal D'.$ We can use $\alpha$ to categorize the labelers as follows: 
 \begin{list}{$\bullet$}{} 
\item {\em Random guesser} is the  labeler with  $\alpha$ close to zero. This labeler is either a lazy labeler who randomly assigns the labels without paying any attention to the instances or an inexperienced labeler.
\item {\em Good labeler} is the labeler with $\alpha$ close to 1. He does a good job of labeling.
\item {\em Malicious/Adversarial labeler} is the labeler with  $\alpha$ close to -1. He guesses the correct label and then flips it. 
 \end{list}

\subsubsection{\bf Instance Difficulty}
 
 Similarly, the difficulty of an instance is defined as:
\begin{equation}\label{beta}\beta_i=(p'_i-q'_i)(1-E'_i)+1\end{equation}

where $\;\;\;\;\;\;\;\;p'_i=\frac{\;M_i^+}{M}\;\;\;\;\;q'_i=1- p'_i$,  $p'_i$ is the probability of getting a correct label for instance $i$, from the crowd labeler and  $M_i^+$ is the number of correct labels given to the instance $i$. Also,
  \begin{equation}\;\;\;\;\;\; E'_i= - p'_i log(p'_i)-q'_i log(q'_i)\;\;\; \end{equation}
 represents the entropy for the instance $i$ which measures the uncertainty in our assessment of the difficulty of the instance. All these values are calculated using the expert labeled instances. 
 We have added 1 to the formula in \eqref{beta} because we find it more convenient mathematically to make the value of $\beta$ positive. 
Another reason for adding 1 is that we cannot assume the difficulty level to be negative, just because the labelers did a bad job of labeling. We have  $\beta_i \in (0,2)$ which is used to categorize the instances as follows:
 \begin{list}{$\bullet$}{} 
\item {\em Easy instance} is the one with  $\beta$ close to 2. 
 \item {\em Average difficulty instance} is the instance with $\beta$ around 1.
 \item {\em Difficult instance} is the instance with  $\beta$ close to 0.
 \end{list} 

To judge the difficulty level of the remaining $(N-n)$  instances, we define {\em hypothesized labels} $W_i$ as: 
 \begin{equation} \;\;\;\;\;\; W_i= sign(\sum_{j=1}^M\alpha_j L_{ij})  \end{equation}


The rest of $\beta$'s are estimated by: 
\begin{equation}\label{beta2}\;\beta_i=(p''_i-q''_i)(1-E''_i)+1\end{equation}
 where $p''_i, q''_i, E''_i$ are calculated using the hypothesized labels.\\

\subsubsection{\bf Label Aggregation} 
The parameters $\alpha, \beta$ are used to aggregate the labels. As a first step for this aggregation, we calculate the probability of getting a correct label for instance $i$ from the labeler $j$ defined as 
  \begin{equation} P(T_i=L_{ij}| \alpha_j, \beta_i)= \sigma(c\alpha_j \beta_i), \end{equation}
where $T_i$ is the true but unknown label for the instance $i$. In this function, c is a scaling factor with value 3. The reason for multiplying with this scaling factor is to span the range of the function to  [0,1], otherwise the values only map to a subinterval of [0,1]. The value 3 is chosen due to the fact that $\alpha_j \beta_i \in [-2,2] $ and  $c\alpha_j \beta_i \in [-6,6]$, the latter choice maps to all values in the interval [0,1].

%
 
%
    Since in this version of ELICE, we are able to identify random and malicious labelers separately, we can make use of this information. We have incorporated this aspect of knowledge in the aggregation formula.
\begin{equation} A_i= sign(\sum_{j=1}^M\sigma (|c\alpha_j \beta_i|)*L_{ij}* sign(\alpha_j \beta_i)) \end{equation}
    This formula flips the label when the product $\alpha \beta$  is negative, which means $\alpha$ is negative (as $\beta$ is always positive) and the labeler is on the malicious side. If the product  $|\alpha \beta|$ has large value, logistic function will weight the label higher and for small value of $|\alpha \beta|$ the weight is small. So for a given instance, when the labeler is random the weight assigned to the label will be low, when the labeler is good or malicious the weight is high. But for the malicious labeler, label is automatically flipped because of being multiplied to  $sign(\alpha \beta)$. This case is specially helpful when many labelers are malicious.

\subsection{ELICE 2 with Clustering}
ELICE 2 also has a cluster-based variant. We cluster the data  and choose equal number of instances from each cluster, to get expert-labels. The rest of the method remains the same.

\section{ELICE 3 Framework}
\label{elice_plus1}

%
In the previous versions of ELICE, we have assumed the availability of domain experts who provide correct labels without making mistakes. Therefore, expert-labeled instances serve as ground truth. But sometimes ground truth is not known for certain due to one or more of the following problems in the crowd-labeling scenario:
  \begin{list}{$\bullet$}{} 
 \item Expert-labels can be wrong due to the complexity of the task.
 \item Experts do not agree on one label and have diverse opinion.
 \item  A ground truth cannot be obtained using methods other than expert-evaluation or has a high acquisition cost (e.g., biopsy in the case of a brain tumor.)
 \end{list}
 In this situation, we propose to add more information other than expert-labeled instances by involving labeler to labeler and instance to instance comparisons. Since the expert labels are available for a subset of instances and have a chance of being wrong, incorporating crowd labels, which are available for the whole dataset can help. This can increase the chance of refining the estimates of the labeler ability and instance difficulty.
 
  In this version of ELICE, the initial inputs of $\alpha$ and $\beta$ are taken from ELICE 2 with the only difference that the expert labels are not necessarily ground truth. Based on this information the pairwise comparison is performed. While this version of ELICE is computationally more expensive than ELICE 1 and 2, it can be helpful when ground truth is not known with certainty. To reduce the computational complexity, we also propose ELICE with circular comparison.

\subsection{ELICE 3 with Pairwise Comparison}

 In this variant of ELICE, we use a generalization of the  model in \cite{Bradley1952,Huang}. In this generalized model, pairwise comparison is used to rank teams of players of a game based on their abilities. The approach uses the previous performance of the players as an input to the model. 
We use a similar idea to find the expertise of the labelers and difficulty of the instances.

We obtain the average score of the labelers and instances that is calculated using $\alpha$'s and $\beta$'s, which we get through the expert evaluation. In our approach we compare labeler to labeler and instance to instance.  There are $M^{'}={M \choose 2}$ pairwise comparisons for $M$ labelers  and $N^{'}={N \choose 2}$ pairwise for $ N$ instances.  
The level of ability of a labeler $j$ based on pairwise comparison is denoted by $\alpha ^\prime_j$. Similarly, the difficulty of instance $i$ based on the pairwise comparison is denoted by $\beta ^ \prime _i$.


  The procedure for  finding $\alpha^\prime_j$'s is described as follows.  
  We assume that the actual performance of labeler $j$, which is represented by a random variable $ X_j$, has some unknown distribution. In order to avoid computational difficulties we assume that the  $ X_j$  has a doubly exponential extreme value distribution with a mode equal to   ${\alpha^\prime_j}$. 
                \begin{equation}P(X_j \leq x)=exp(exp{-(x-\alpha^ \prime_j)})  \end{equation}
  
  
   This distribution ensures that the extreme values are taken into consideration, and variance is directly affected by the values but is not dependent on the mean of the distribution. Hence according to  \cite{Huang}:
      
      \begin{equation} \label{eq1} P(C_j ~\mbox{is more expert  than} ~C_k ) =\frac{exp(\alpha^\prime_j)}{(exp(\alpha^\prime_j)+exp(\alpha^\prime_k))} \end{equation}
          
           where $C_j$ is the crowd labeler $j$ and $C_k$ is the crowd labeler $k$.
         We use $\alpha$'s and $\beta$'s to calculate the average score of reliability of the labelers  denoted by $P_j$.
                    \begin{equation} \mbox{P}_j=\frac{1}{M} \sum_{i=1}^N \sigma(c\alpha_j \beta_i) \end{equation}

 The average score is calculated to make sure that, while doing a pairwise comparison of labelers, their average performance on the whole dataset is taken into consideration.
We assume that the probability of one labeler being better than another labeler is estimated by the ratio of the average score of the labelers \cite{Huang}. This can be expressed in the form of an equation by using equation \ref{eq1} and the ratio of ${\mbox{P}_j}$ and ${\mbox{P}_j+ \mbox{P}_k}$.       
       \begin{equation} \frac{exp(\alpha^\prime_j)}{(exp(\alpha^\prime_j)+exp(\alpha^\prime_k))} \approx\ \frac{\mbox{P}_j}{\mbox{P}_j+ \mbox{P}_k}  \end{equation}
           $$ \implies      \frac{1}{(1+exp(-(\alpha^\prime_j-\alpha^\prime_k)))} \approx\ \frac{1}{1+\frac{ \mbox{P}_k}{\mbox{P}_j}}$$            
                                         \begin{equation} \implies (\alpha^\prime_j-\alpha^\prime_k) \approx\  log(\frac{ \mbox{P}_j}{ \mbox{P}_k}) \end{equation}
                       
                       This can be formulated as the least square model:                      
                      \begin{equation} \implies   \underset{ \boldsymbol \alpha^\prime} {min}\;\;\;\;  \overset  {M}{  \underset {j=1,k=j+1}{\sum}} [(\alpha^\prime_j-\alpha^\prime_k) -{log(\frac{ \mbox{P}_j}{ \mbox{P}_k})})]^2  \end{equation}
                                          
              Which can be written in the matrix form as
                           \begin{equation} \underset{ \boldsymbol \alpha^\prime}{min} \;\;\;\; ({\bf{ G \boldsymbol \alpha^\prime-d}})^T ({\bf{ G \boldsymbol \alpha^\prime-d}})  \end{equation}

                      ${ \boldsymbol G}$ is a matrix of order ${M^\prime}$ x $M$. The rows represent comparisons and columns  represent the labelers. The matrix is defined as  
                                          
                     \begin{equation}G_{lj}=\begin{cases} \;\;1  \;\;\;\; \mbox{ $j$ is the first labeler in  the $l^{th}$ comparison} \\  -1 \;\;\;\; \mbox{$j$ is the second labeler in  the $l^{th}$ comparison}\\  \;\;\; 0 \;\;\;\;\mbox{labeler $j$ is not in  the $l^{th}$ comparison}\end{cases}  \end{equation}
                      where
                       $\;\;\;\;\; j=1,2, \dots, M; \;\;\;\;l= 1,2,\ldots,M^{'}$ \\
                       Also,   \begin{equation}\;\;\;\;\;\;\;  d_{(j,k)}=log(\frac{ \mbox{P}_j}{ \mbox{P}_k})\;\;\;\;\; \end{equation}
                        where  $\;\;\;\;\;\;j=1,2,\ldots, M; \;\; k= j+1, j+2, j+3,\ldots,M$. \\
          
          We can derive the following expression:
                \begin{equation} \boldsymbol \alpha^\prime=\bf{ (G^T G)^{-1} G^Td}  \end{equation}
                    
                In order to avoid the difficulties when the matrix $G^TG$ is not invertible, we add a regularized term $ \mu { \boldsymbol  \alpha^\prime}^T { \boldsymbol  \alpha^\prime}$ where $\mu $ is a very small real number which can be learned heuristically.
 
       \begin{equation}  \underset{ \boldsymbol \alpha^\prime}{min} \;\;\;\; ({\bf{ G \boldsymbol {\alpha^\prime}-d}})^T ({\bf{ G \boldsymbol {\alpha^\prime}-d}})+\mu\boldsymbol {\alpha^\prime} ^T \boldsymbol {\alpha^\prime}  \end{equation}
                        
                    The resulting expression for  ${\boldsymbol {\alpha^\prime}}$ we get is,
                     \begin{equation}{\boldsymbol {\alpha^\prime}}=\bf{ (G^T G+\mu I)^{-1} G^Td}   \end{equation}
                      where ${\boldsymbol I}$ is the identity matrix.
                      
This procedure can be repeated to find an expression for $  \boldsymbol {\beta^\prime} $'s.
First we find the average score of the difficulty of each instance:
                     \begin{equation} \mbox{Q}_i=\frac{1}{N}\overset {M} { \underset {j=1}{\sum}} \sigma(c\alpha_j \beta_i) \end{equation}        

                    Then repeating the above mentioned steps and adding 1's to make $\beta ^{\prime}$'s positive, we get
                    \begin{equation} \boldsymbol {\beta^\prime}=\bf{ (H^T H+\nu I)^{-1} H^Td^\prime} +1  \end{equation}
    where    $\;\;{ d^\prime_{(i,p)}}=log(\frac{\mbox{Q}_i}{\mbox{Q}_p})\;$ and  $\;\;\;i=1,2, \ldots, N; \;\;\;\;\; p= i+1, i+2, \ldots, N.$

Also,  \begin{equation}H_{ri}=\begin{cases} \;\;1  \;\;\;\;  \mbox{$i$ is the first instance in the $r^{th}$ comparison  } \\  -1 \;\;\; \mbox{$i$ is the second instance in  the $r^{th}$ comparison }\\  \;\;0 \;\;\;\;\mbox{instance $i$ is not in  the $r^{th}$comparison }\end{cases} \end{equation}
     such that $\;\;\;\;\;\;\;\; i=1,2, \ldots,N$;  $r= 1,2,3...N^{'}$.\\
     
     After finding the $ {\alpha^\prime}_j$'s and $ {\beta^\prime}_i$'s we use them to infer the labels.
         \begin{equation}A_i= sign(\sum_{j=1}^M\sigma (|c\alpha_j \beta_i|)*L_{ij}* sign(\alpha_j \beta_i))   \end{equation}



 As in the previous version of ELICE, we multiply $\alpha^\prime_j \beta^\prime_i$ by a scaling factor $c$ to make sure that the range of the values is mapped to the whole range of the logistic function i.e., [0,1] and not just on its subinterval. This also serves to make the difference between the expertise of  workers on different instances more pronounced. Since in this case the value of the product $|\alpha_j \beta_i| <<1$, the value of $c$ has to be large. We used $c=100$, chosen heuristically through experiments.

\subsection{ELICE 3 with Circular Comparison} 
ELICE with circular comparison is a variant of ELICE with pairwise comparison. Instead of making comparison of every two labelers, it compares labelers to labelers and instances to instances in a circular fashion, for example, $1$ to $2$, $2$ to $3$, $\ldots$, $i$ to $i+1$, $\ldots$, $M-1$ to $M$, $M$ to $1$. Our empirical results show that this produces results as good as  ELICE with pairwise comparison but substantially reduces the computational cost.




\section{Empirical Evaluation}
\label{experiments}
\vspace*{0.3cm}

We implemented ELICE
and its variants in Matlab. 
We compare our method to
Majority voting, GLAD and GLAD with clamping \cite{WhitehillNIPS2009}, Dawid and Skene \cite{Dawid:Skene:79}, EM (Expectation Maximization), Karger's iterative method \cite{Karger}, Mean Field algorithm and Belief Propagation \cite{Qiang}. Please note that Karger's iterative method, Mean Field method and Belief Propagation have two versions each due to different parameter setting. All of these methods were also implemented in MATLAB and in most cases the code was obtained from authors of the methods. We conducted the experiments on the following datasets:
\begin{list}{$\bullet$}{} 
\item Five datasets from the UCI repository \cite{AsuncionNewman:2007}: IRIS, Breast Cancer, Tic-Tac-Toe, Chess, Mushroom (section \ref{uci}).  Crowd labels are simulated for different rates of random/malicious crowd labelers in the pool of labelers.
\item Two real applications Tumor Identification dataset, Race Recognition dataset (section \ref{cancer} and \ref{race_recognition}) for which we use MTurk to acquire labels from the crowd. 
\end{list}

\begin{figure}[htbp]
\begin{center}
\vspace*{-6cm}
\includegraphics[scale=0.75]{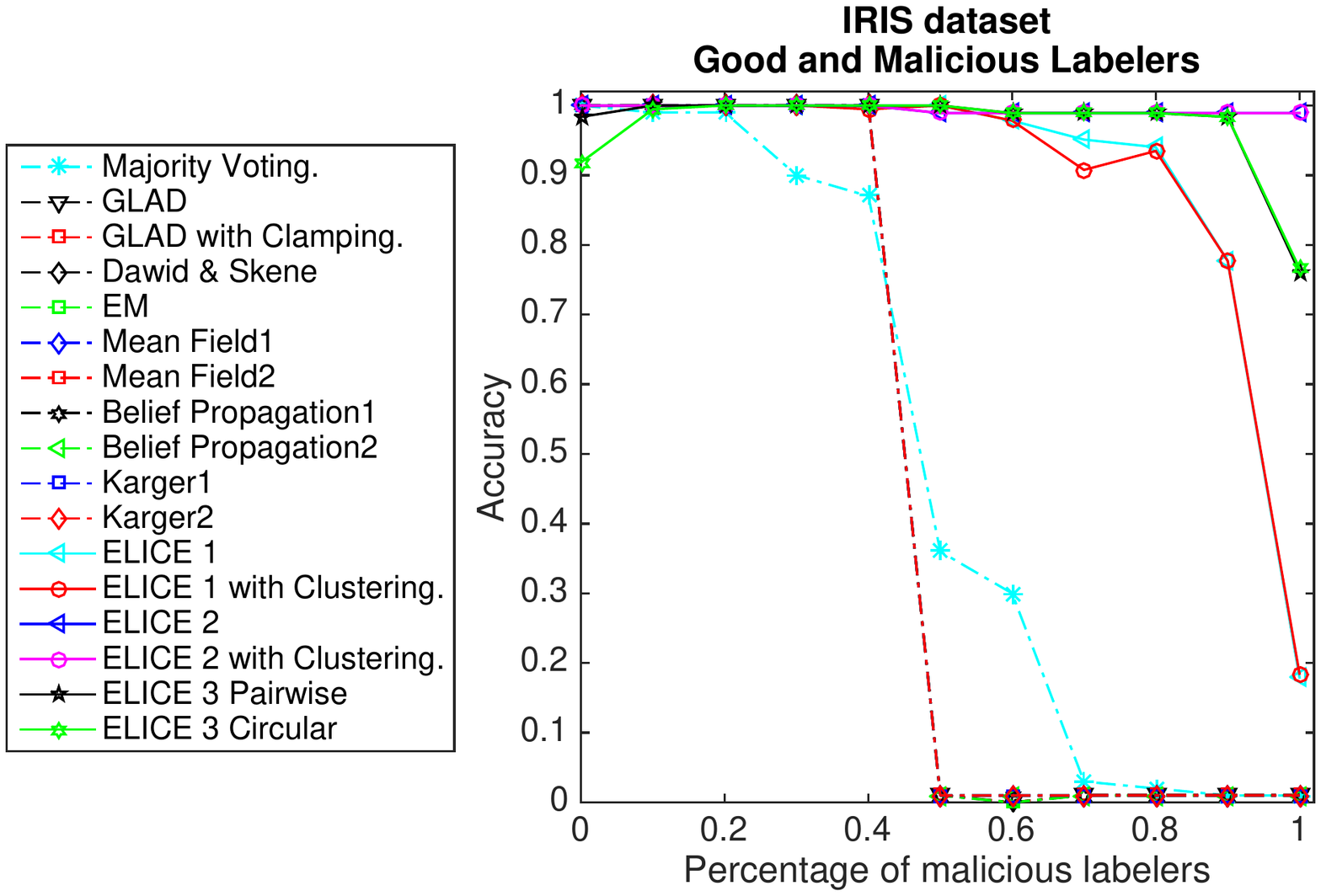} \\
\vspace*{-11cm}
\includegraphics[scale=0.75]{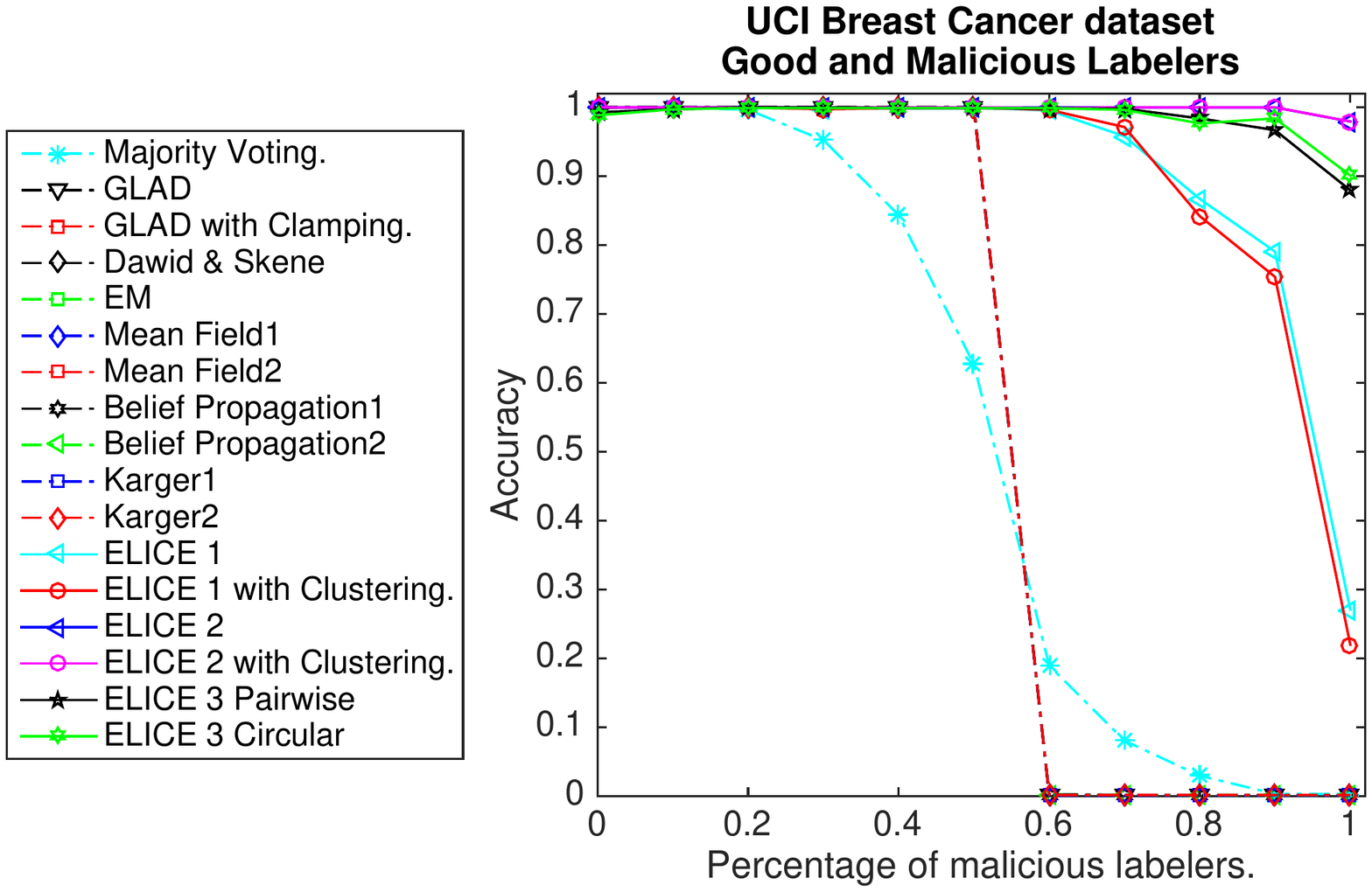}
\vspace*{-6cm}
\end{center}
\caption{(Top) IRIS dataset. (Bottom) UCI Breast Cancer dataset. Good and malicious labelers. }
 \label{GM1}
 \end{figure}

\begin{figure}[htbp]
\begin{center}
\vspace*{-6cm}
\includegraphics[scale=0.73]{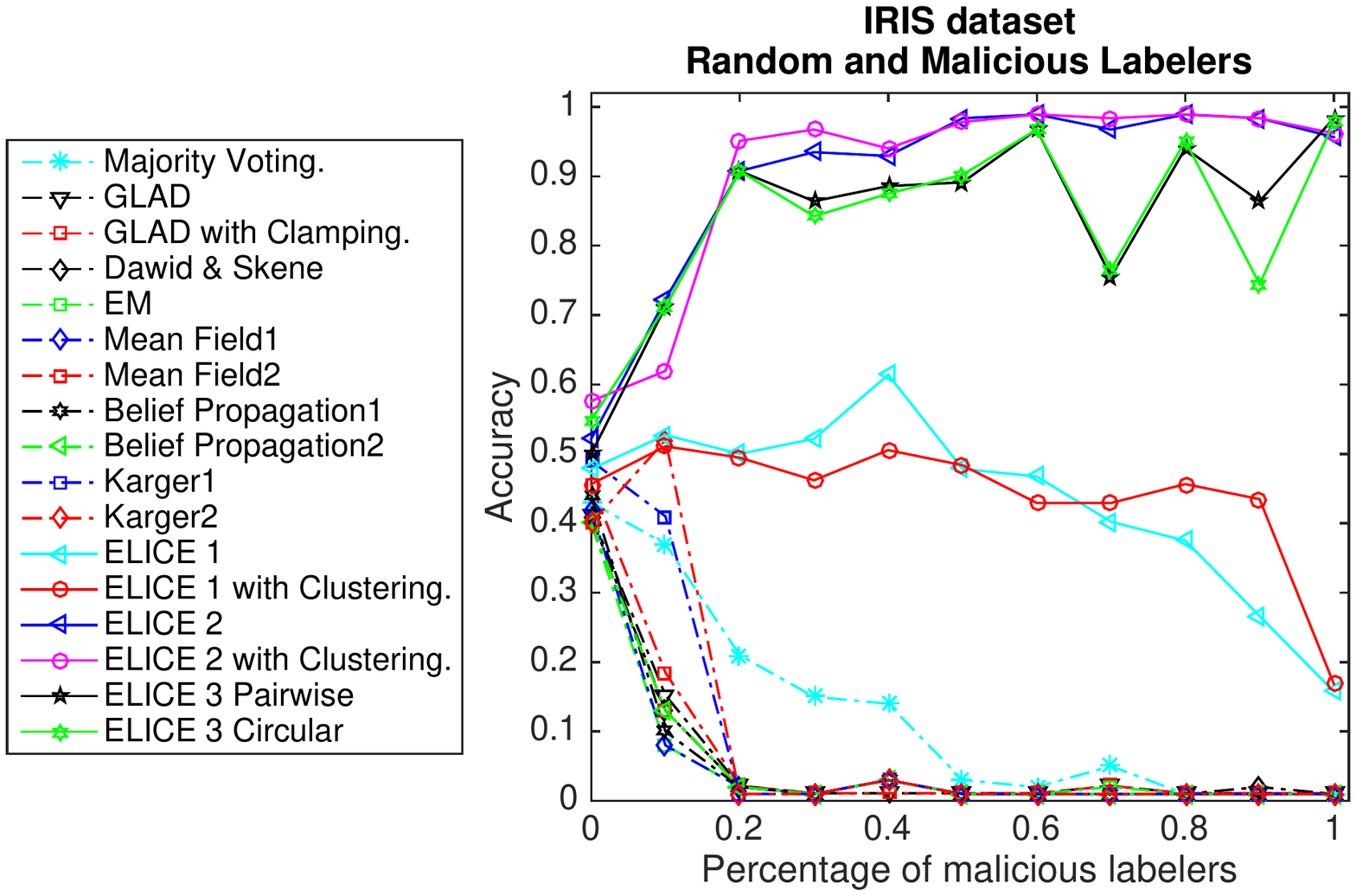} \\
\vspace*{-11cm}
\includegraphics[scale=0.75]{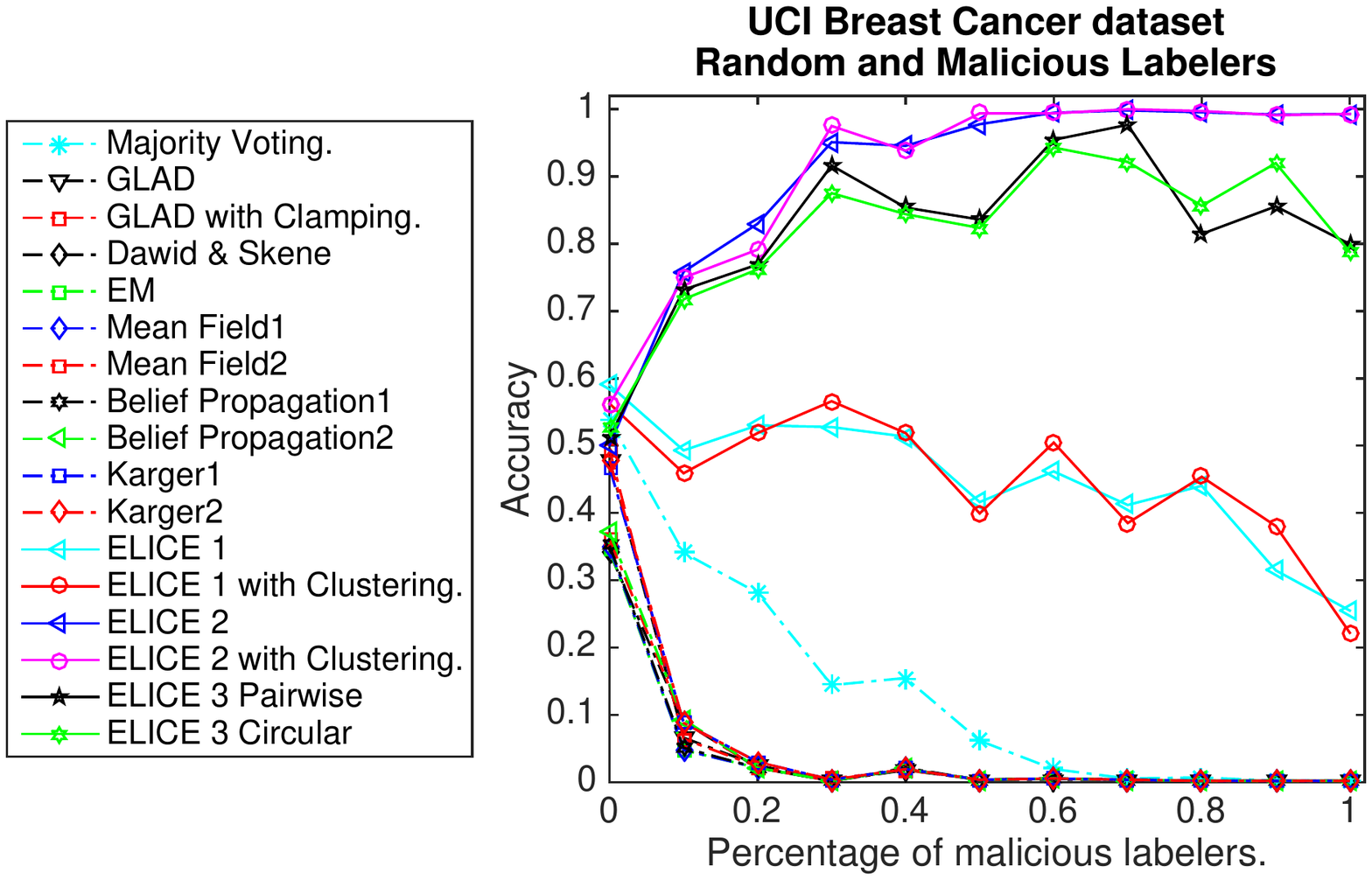}
\vspace*{-6cm}
\end{center}
\caption{(Top) IRIS dataset. (Bottom) UCI Breast Cancer dataset. Random and malicious labelers.}
 \label{RM1}
 \end{figure}

\begin{figure}[htbp]
\begin{center}
\vspace*{-6cm}
\includegraphics[scale=0.73]{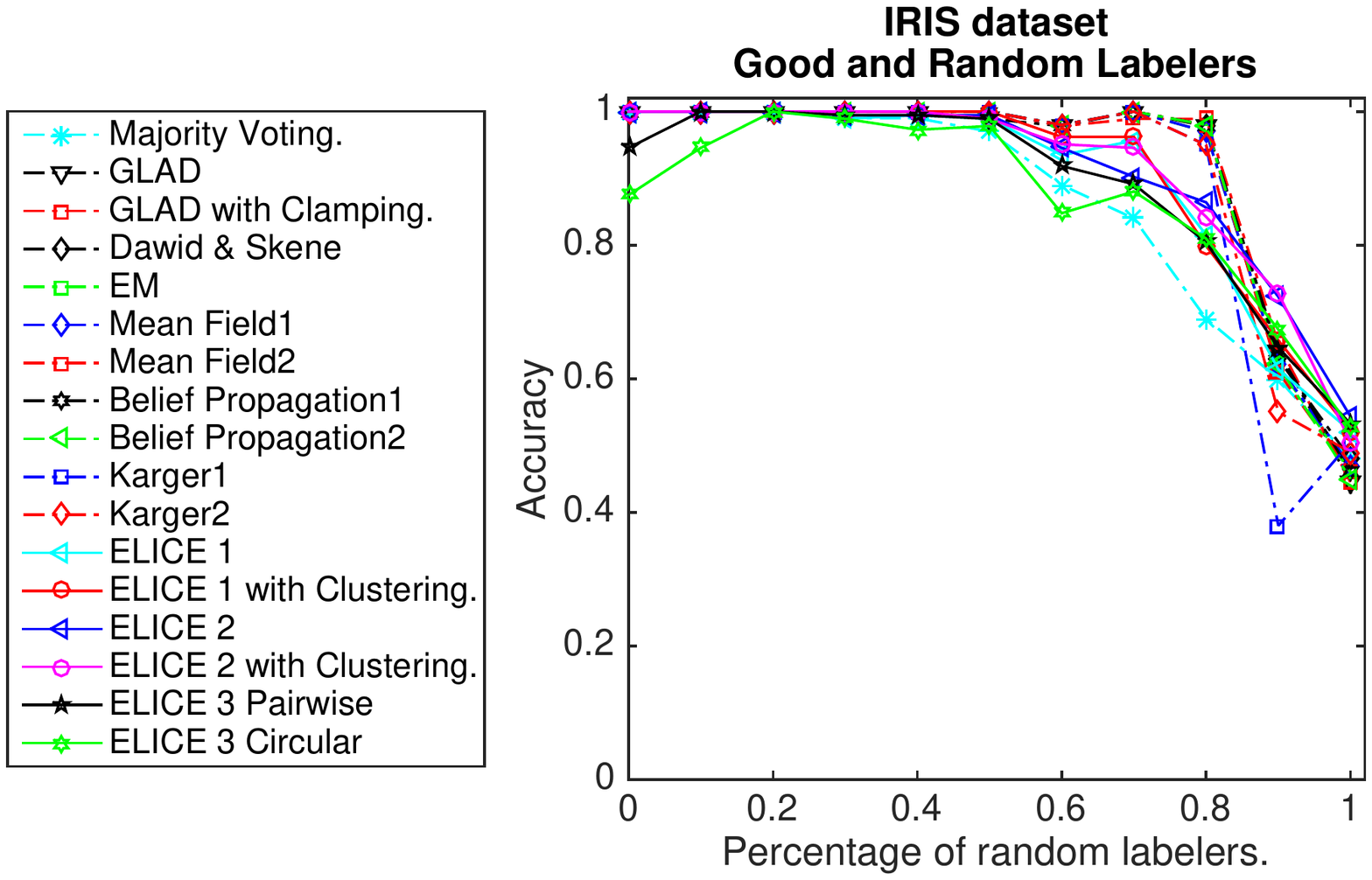} \\
\vspace*{-11cm}
\includegraphics[scale=0.75]{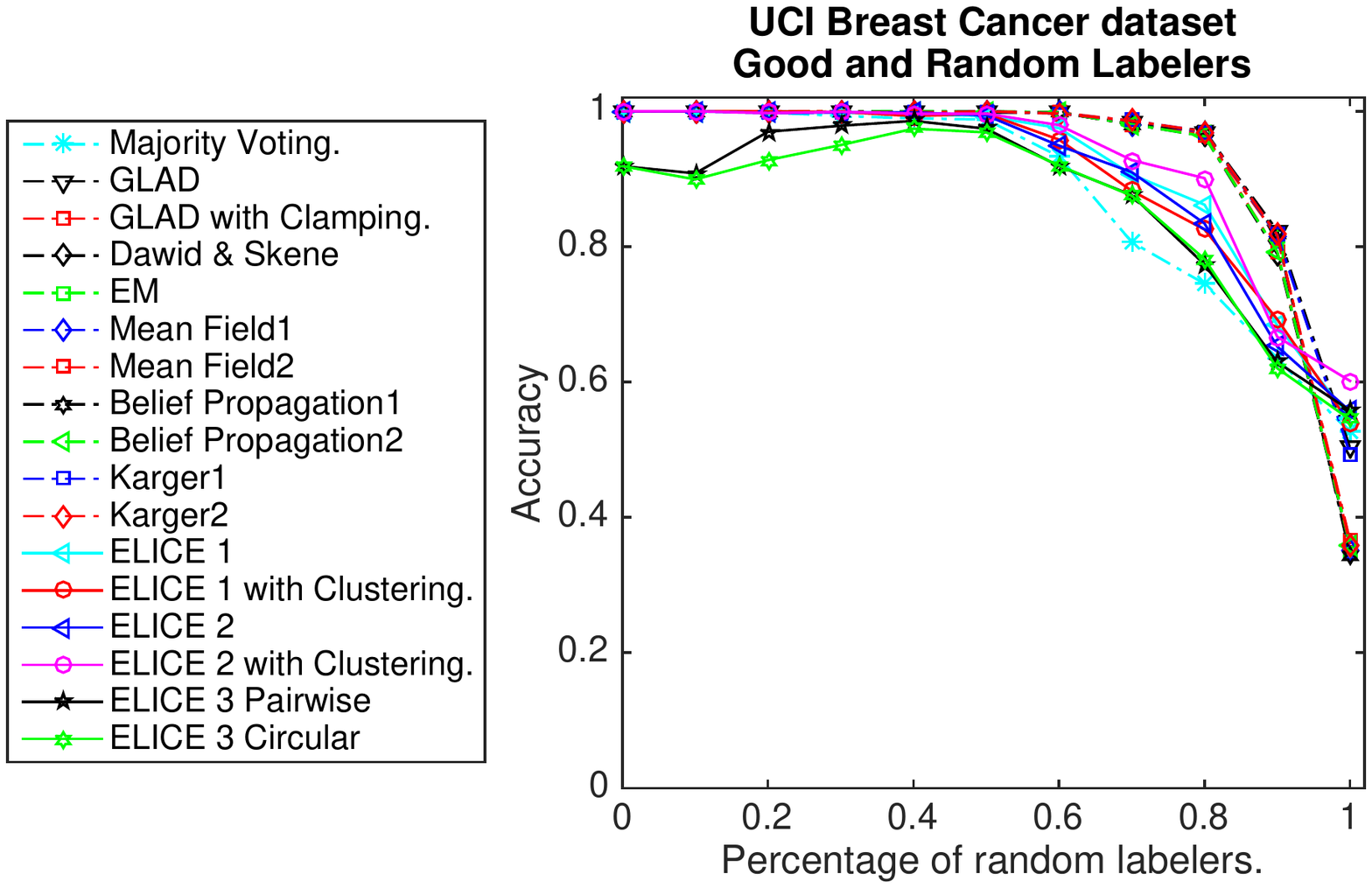}
\vspace*{-6cm}
\end{center}
\caption{(Top) IRIS dataset. (Bottom) UCI Breast Cancer dataset. Good and random labelers. }
 \label{GR1}
 \end{figure}

 \subsection{ UCI datasets }
 \label{uci}
In this experiment, we selected five datasets from the UCI repository: Mushroom, Chess, Tic-Tac-toe, Breast cancer and IRIS (with this latter restricted to 2 classes by excluding the instances from the rest of the classes). 

 \subsubsection{Experimental Design} We simulated 20 crowd labels for each instance in these experiments. The labels were generated so that a good crowd labeler makes less than 35\% mistakes, a random crowd labeler makes 35\% to 65\% mistakes and a malicious crowd labeler makes 65\% to 100\% mistakes.
These were created by inverting $x$\% of the original labels in the dataset, where $x$ is a random number between 0 and 35 for good labeler, 35 to 65 for random labeler and 65 to 100 for malicious labeler.
We randomly selected $n$ number of instances to play the role of the expert-labeled instances.\\

\begin{table*}[htbp]

{\scriptsize  
\hspace*{-1cm}\begin{tabular}{l l c c c c c c c}
 \\ \hline

\multirow{3}{*}{\scriptsize Random and/or Malicious Labelers }
 &Dataset ($\cal D$)  & Mushroom & Chess &Tic-Tac-Toe  & Breast Cancer & IRIS  \\  \cline{2-7}
 &Total instances ($N$)& 8124 & 3196& 958  &569 &100  \\    \cline{2-7}
 &+ve/-ve instances  & 3916/ 4208 &1669/1527 &626/332 &357/212 &50/50\\   \cline{2-7}
   &Expert labels($n$) &20 &8 &8 &8 &4  \\   
        \hline
  
 \multirow{17}{*}{Less than 30\%  }& Majority Voting  & 0.9913 & 0.9822 & 0.9851     & 0.9978 & 0.9950\\ 

   &GLAD &  {\bf 1.0000} & {\bf 1.0000}  & {\bf 1.0000}                        & {\bf 1.0000} & {\bf 1.0000}         \\ 

    &GLAD with clamping & {\bf 1.0000} & {\bf 1.0000} & {\bf 1.0000}                 & {\bf 1.0000} & {\bf 1.0000}   \\

     &Dawid Skene          & {\bf 1.0000}       &{\bf 1.0000}          & {\bf 1.0000}           &  {\bf 1.0000} & {\bf 1.0000}\\
     &EM          &{\bf 1.0000}        &{\bf 1.0000}           & {\bf 1.0000}           &  {\bf 1.0000} & {\bf 1.0000}\\
     &Belief Propagation 1  & $\dagger$    &0.9918                 & {\bf 1.0000}            &  {\bf 1.0000} & {\bf 1.0000}\\
     &Belief Propagation 2   & $\dagger$      &$\dagger$                   & {\bf 1.0000}          &  {\bf 1.0000} & {\bf 1.0000}\\
     &Mean Field 1              &{\bf 1.0000}       &{\bf 1.0000}             & {\bf 1.0000}         &  {\bf 1.0000} & {\bf 1.0000}\\
     &Mean Field 2             &{\bf 1.0000}        &{\bf 1.0000}             & {\bf 1.0000}         &  {\bf 1.0000} & {\bf 1.0000}\\
     &Karger 1                    &{\bf 1.0000}        &{\bf 1.0000}            & {\bf 1.0000}          &  {\bf 1.0000} & {\bf 1.0000}\\
     &Karger 2                    & {\bf 1.0000}      &{\bf 1.0000}           &{\bf 1.0000}            &  {\bf 1.0000} & {\bf 1.0000}\\

       &ELICE 1  &0.9988 & 0.9994 &0.9989                       & 0.9993 & {\bf 1.0000} \\ 
 
     &ELICE 1 with clustering & 0.9993 & 0.9994  &0.9989         & 0.9991 &{\bf 1.0000}\\ 
     
     &ELICE 2  & 0.9997 & 0.9999&{\bf 1.0000}                &0.9989 & {\bf 1.0000} \\ 
 
     &ELICE 2 with clustering &  0.9998  & {\bf 1.0000}& {\bf 1.0000}           &0.9991 &{\bf 1.0000}\\ 
     &ELICE 3 Pairwise &*  & 0.9768  &  0.9925                     &     0.9701 &     0.9959 \\ 
  &ELICE 3 Circular & 0.9567  &0.9800 &    0.9842          &    0.9635  &        0.9891\\
        \hline

   \multirow{17}{*}{30\% to 70\%}&Majority Voting  &0.6587   &0.6541       &0.6654  &       0.7179 &     0.6900\\ 

     &GLAD                         &  0.7494 &   0.7502       &0.7503 & 0.7504    &    0.7473\\ 
     &GLAD with clamping  &  0.7494 & 0.7501  &0.7505 & 0.7504  &     0.7473 \\ 
    
     &Dawid Skene          & 0.5001   &0.7498   & 0.5003 &  0.7504  &     0.7475\\
     &EM          &0.5001   &0.7498   & 0.5003 &  0.7504 &     0.7475\\
     &Belief Propagation 1   & $\dagger$  &0.7107   & 0.5003 & 0.5004 &     0.7500\\
     &Belief Propagation 2   &$\dagger$ &$\dagger$   & 0.5003 & 0.7504  &     0.7525\\
     &Mean Field 1              &0.5002   &0.7498   & 0.5003 &  0.7504  &         0.7500\\
     &Mean Field 2              &0.5001   &0.7498   & 0.5003 & 0.7504  &         0.7525\\
     &Karger 1                     &0.5002   &0.7498   & 0.5005 &   0.6254  &     0.7525\\
     &Karger 2                     &0.5003   &0.7498   & 0.5005 &  0.7504  &     0.7525\\

     &ELICE 1  &0.9779 & {\bf 0.9981} & 0.9915                       &0.9701  &0.9837 \\ 
      &ELICE 1 with clustering &  0.9731 &0.9677 & 0.9839      &0.9650  & 0.9715\\ 
      &ELICE 2                         &  0.9975 & 0.9964 &{\bf 0.9991}  &0.9973  & 0.9932 \\ 
      &ELICE 2 with clustering &  {\bf 0.9985} &  0.9973 &  0.9987  & {\bf 0.9987 } & {\bf0.9960}\\ 
      &ELICE 3 Pairwise &$\ast$           & 0.9948  & {\bf 0.9991}     &0.9951       & 0.9905 \\ 
      &ELICE 3 Circular & 0.9978       & 0.9907  &  {\bf 0.9991}  &0.9949  &        0.9878\\    
   \hline

 \multirow{17}{*}{More than 70\%}&Majority Voting &  0.0889 & 0.0824  &0.0814 &    0.1060 &    0.0300 \\ 
   &GLAD           &     4.1071e-04    &0.0031            &   0.0011         & 0.0024    &     0.0145  \\ 
    &GLAD with clamping & 4.1071e-04 & 0.0031    &  0.0014          &0.0018     &     0.0145 \\ 
     
     &Dawid Skene            &1.6412e-04    &9.3867e-04   &  0.0045       & 0.0023   &     0.0133\\
     &EM           & 1.6412e-04  &8.3438e-04   & 0.0049         & 0.0023    &     0.0133\\
     &Belief Propagation 1   &$\dagger$ &0.1315          & 0.0049         & 0.0023    &     0.0133\\
     &Belief Propagation 2   & $\dagger$ &$\dagger$         & 0.0045         &0.0023     &     0.0133\\
     &Mean Field 1              &  1.6412e-04 &8.3438e-04    & 0.0049         & 0.0023    &     0.0133\\
     &Mean Field 2              &1.6412e-04   &9.3867e-04    & 0.0045         &0.0023     &     0.0133\\
     &Karger 1                     & 3.6928e-04  &0.0021           & 0.0042         & 0.0035     &     0.0100\\
     &Karger 2                     & 3.6928e-04  &0.0021           & 0.0042         & 0.0035     &     0.0100\\

   &ELICE 1                        & 0.7451 & 0.6332 &0.7441&      0.6869 &     0.7065 \\ 
   &ELICE 1 with clustering &  0.7228  &0.6003 & 0.7346  &     0.7020  &     0.6993\\ 
   &ELICE 2                         &0.9900 & 0.9847 &0.9934&      0.9872 &      0.9783 \\ 
   &ELICE 2 with clustering &{\bf 0.9942}   &{\bf 0.9869} & {\bf 0.9956} &{\bf     0.9881}  & {\bf     0.9801}\\ 
   &ELICE 3 Pairwise          &$\ast$  &   0.9848 &   0.9605 &         0.9629 &     0.9656 \\ 
   &ELICE 3 Circular           & 0.9680  & 0.9521  & 0.9590   &    0.9635 &         0.9601\\ 
     
      \hline

  \end{tabular}
}
\caption{\scriptsize{Accuracy of Majority voting, GLAD (with and without clamping) \protect \cite{WhitehillNIPS2009}, Majority voting, Dawid and Skene \cite{Dawid:Skene:79}, EM (Expectation Maximization), Karger's iterative method \protect \cite{Karger}, Mean Field algorithm and BP \protect \cite{Qiang} and ELICE (all versions and variants) for different datasets.  Given results are the  average of 50 runs.  Good labelers: 0-35\% mistakes, Random labelers: 35-65\% mistakes, Malicious labelers: 65-100\% mistakes. \\  $\dagger$ Code for Belief propagation did not converge.\\  $\ast$ Code for ELICE pairwise was parallelized for datasets with more than 3000 instances. But for Mushroom dataset due to high time and space complexity as well as constraints on the available hardware it was not feasible to calculate the results. 
}}
\label{table1}
\end{table*}


\begin{figure}[htbp]
\begin{center}
\vspace*{-5cm}
\includegraphics[scale=0.65]{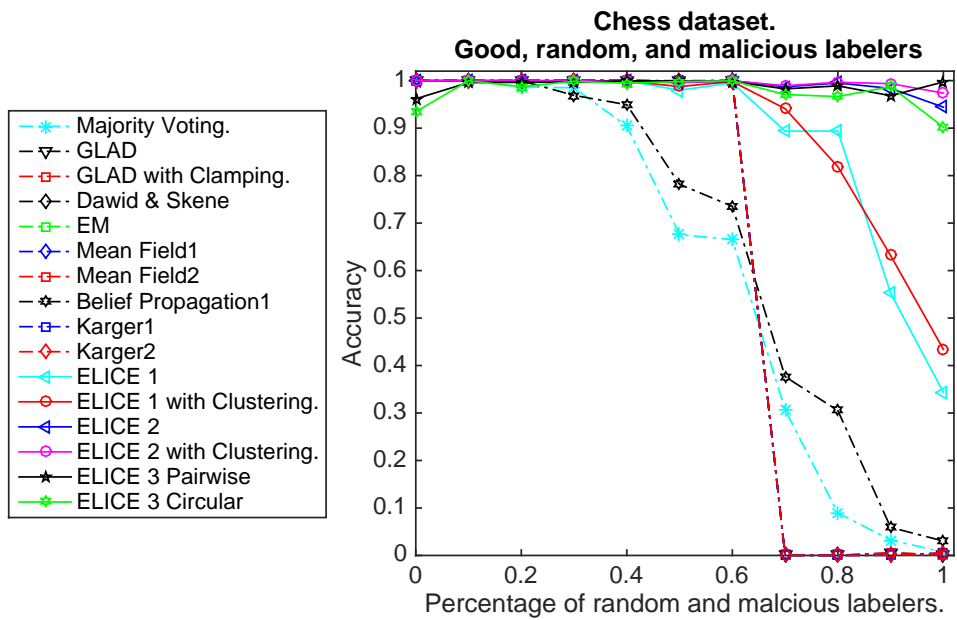}
\vspace*{-5cm}
\end{center}
\caption{ Accuracy vs. percentage of random and malicious labelers. We start with all good labelers and keep on increasing the percentage of random and malicious labelers. Number of expert labels used for  ELICE and all its versions is 20.}
 \label{chess1}
 \end{figure}

\begin{figure}[htbp]
\begin{center}
\vspace*{-6cm}
 \includegraphics[scale=0.7]{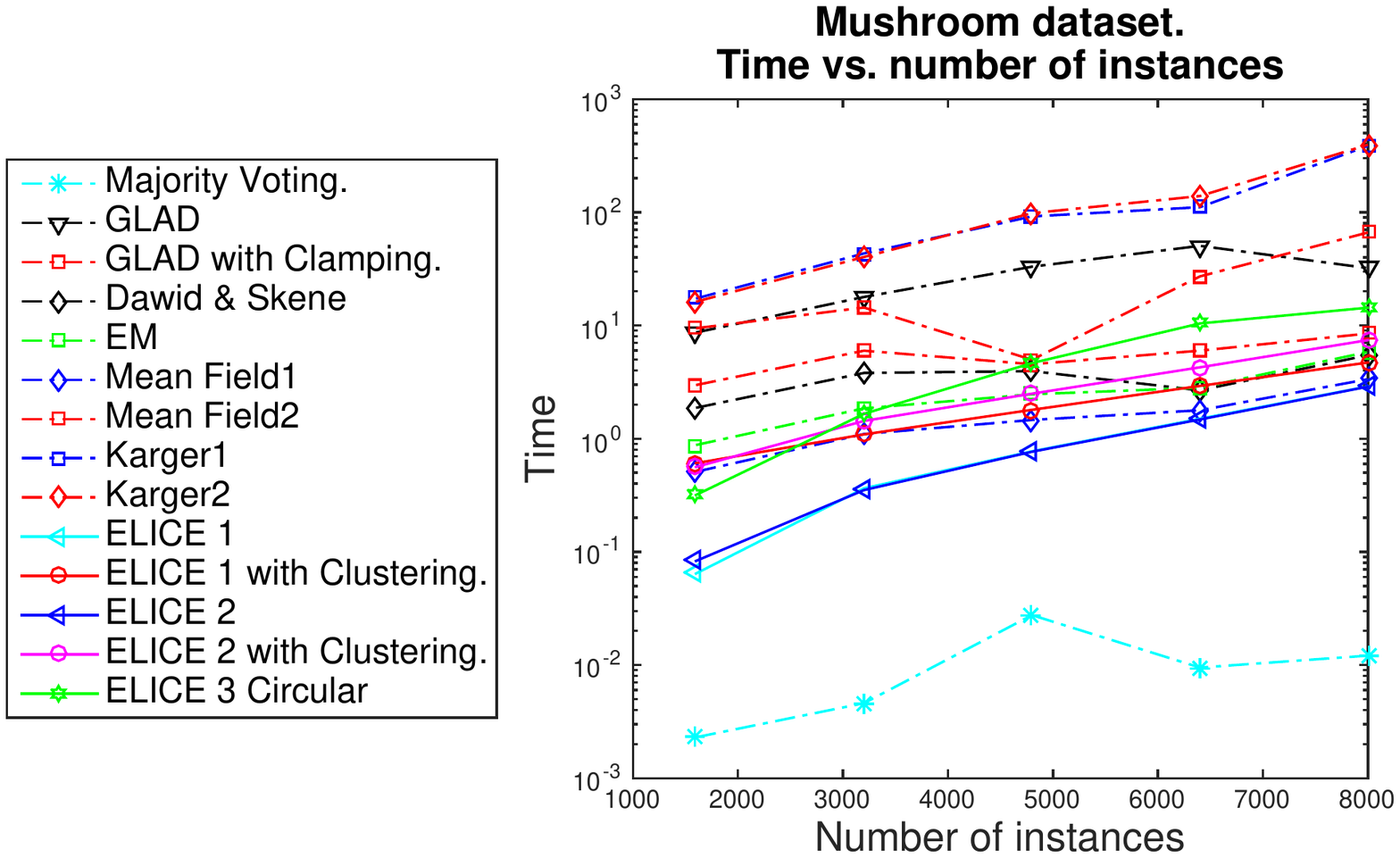}
\vspace*{-6cm}
\end{center}
\caption{ Time vs. Number of instances. Number of expert labels used for  ELICE (all versions and variants) is 20.\\
 {\small Note: Code for Belief Propagation did not converge even after a long time. Code for ELICE pairwise was parallelized for datasets with more than 3000 instances therefore, we do not report its time as it is not comparable to the non-parallelized code.}}
 \label{time1}
 \end{figure}

 \subsubsection{Results} We ran ELICE and its variants, along with the other methods. Table \ref{table1} shows a comparison of accuracy of different methods as compared to ELICE across the five datasets. We use different percentage of bad and malicious labelers while the rest of the labelers are good. Note that all versions of ELICE outperform all other methods even when the percentage of random and/or malicious labelers is increased to more than 70\%. 

 In Figures \ref{GM1}, \ref{GR1} and \ref{RM1}, we show the accuracy of the methodologies for the IRIS and UCI breast cancer dataset for (a) good \& malicious, (b) good \& random and (c) random \& malicious labelers respectively. All these graphs show the superiority of ELICE on other state-of-the-art methods. The experiments also reveal that ELICE is efficient as compared to the other methods. Figure \ref{chess1} shows the accuracy for the Chess dataset, where the labelers are a combination of good, random and malicious. Figure \ref{time1} shows the runtime for Mushroom for all the methods as we increase the number of instances. It should be noted that  for big datasets such as Chess (3196 instances), we used MATLAB's Parallel Computing Toolbox to run ELICE pairwise.

\subsection{Tumor Identification Dataset}
\label{cancer}


%
%


To test our approach on a real-life dataset, we considered a {\em tumor identification dataset}.\footnote{Available on \url{http://marathon.csee.usf.edu/Mammography/Database.html}} 
Early identification of cancer tumor can help in preventing  thousands of deaths but identifying cancer is not an easy task for untrained eyes. 

 \subsubsection{Experimental Design} We posted 100 mammograms on Amazon Mechanical Turk. The task was to identify Malignant versus others (Normal, Benign, Benign without call back.)
The following instruction for appropriate identification was provided to the labelers: 
``A breast tumor is a dense mass and will appear whiter than any tissue around it.  Benign masses usually are round or oval in shape, but a tumor may be partially round, with a spiked or irregular outline as part of its circumference.''

\begin{figure}[h]
\begin{center}
\includegraphics[scale=0.26]{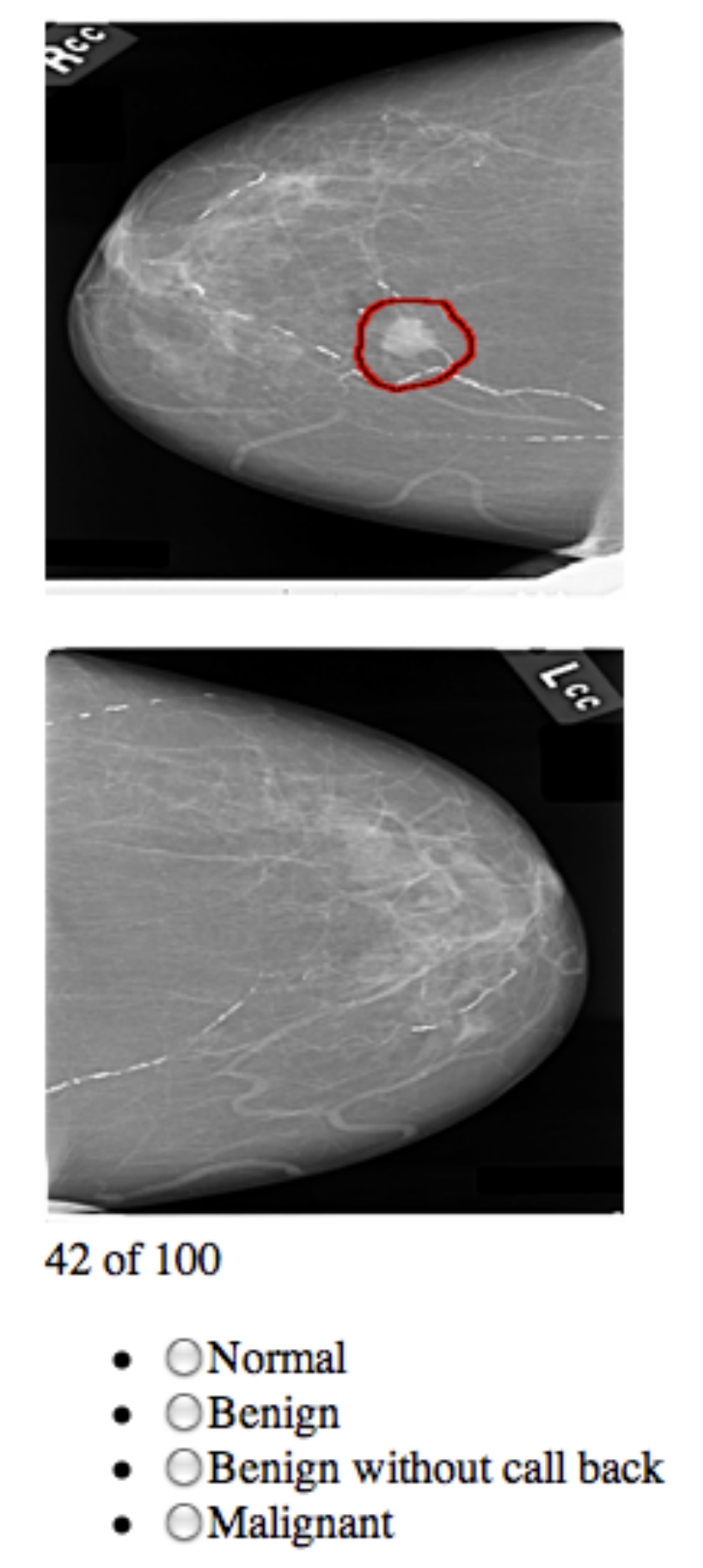}\   \includegraphics[scale=0.26]{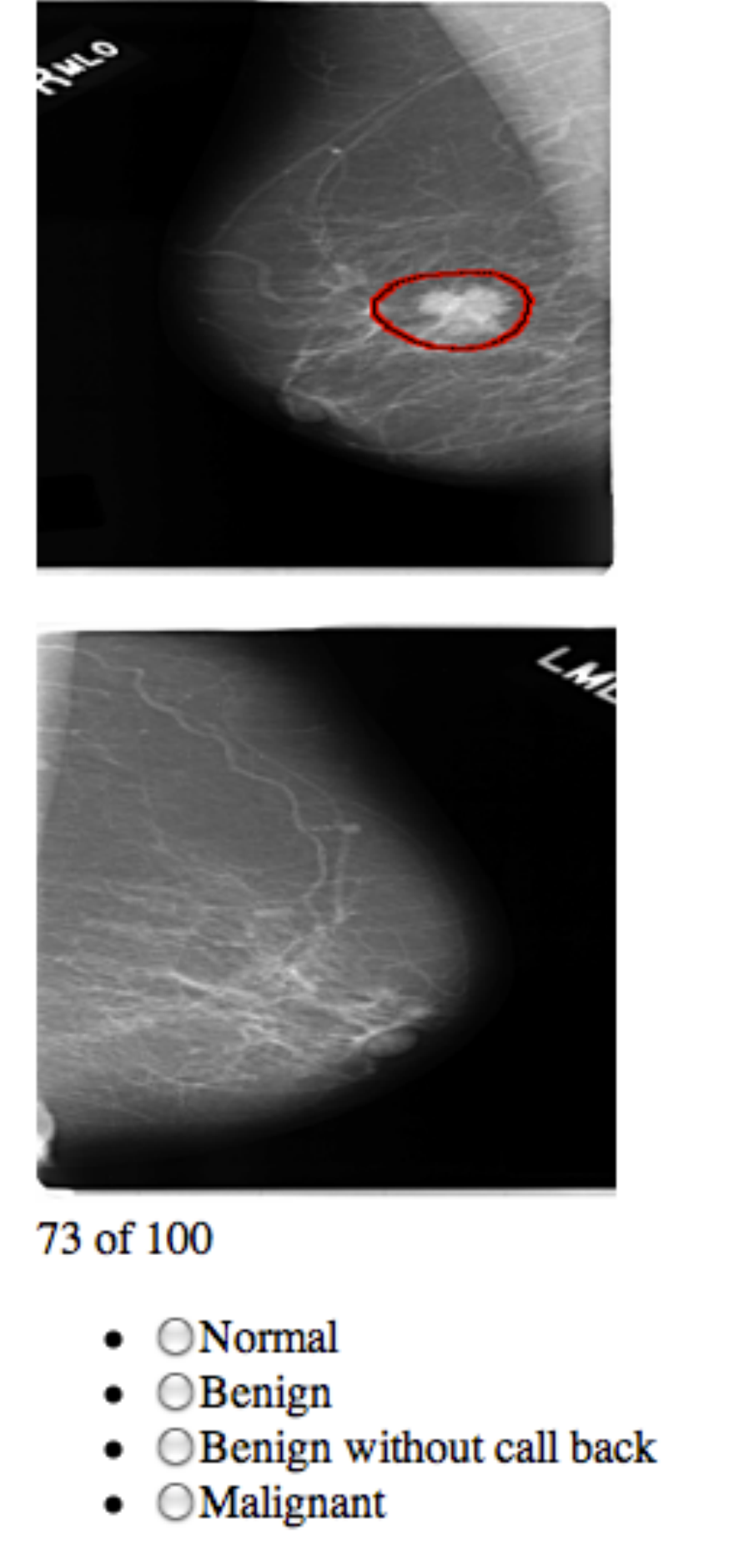}\   \includegraphics[scale=0.26]{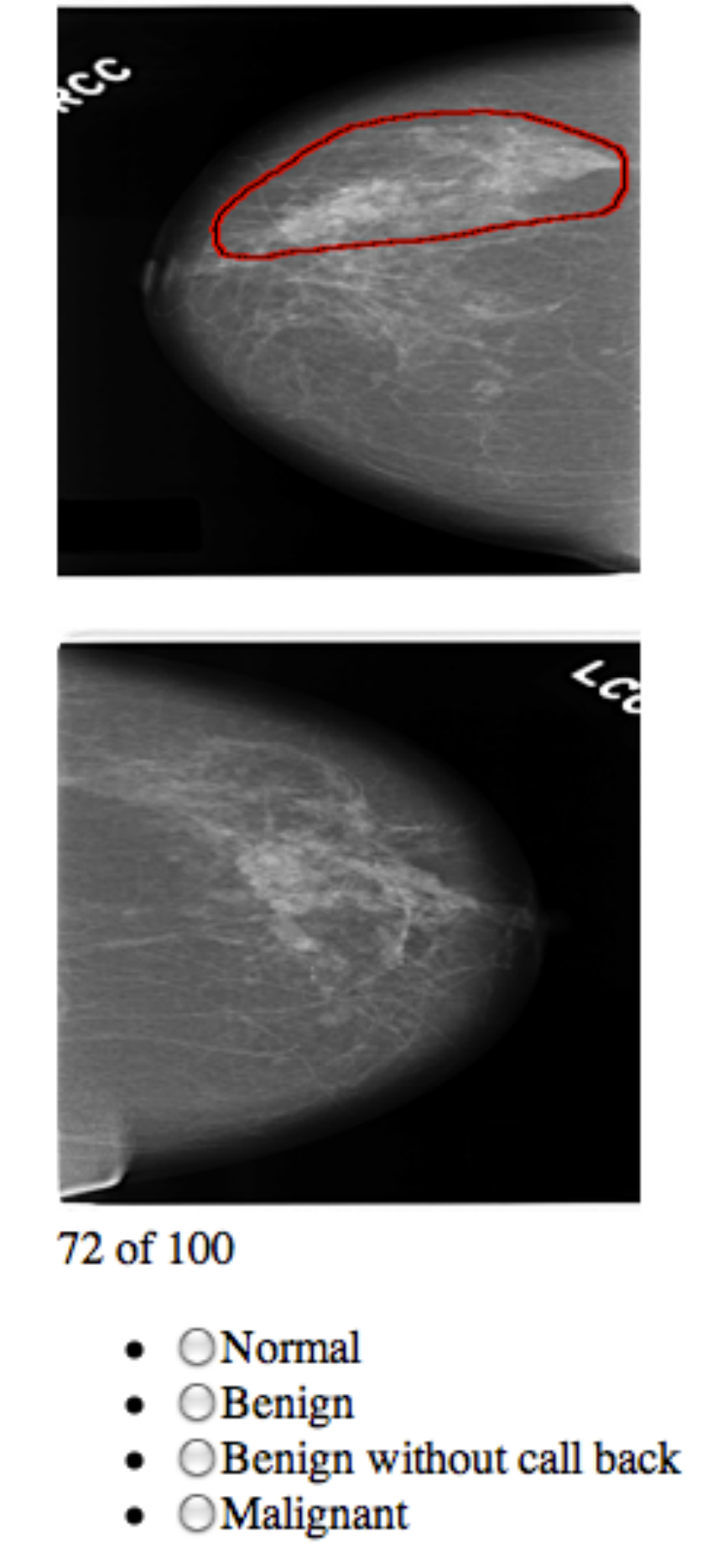}\   \includegraphics[scale=0.26]{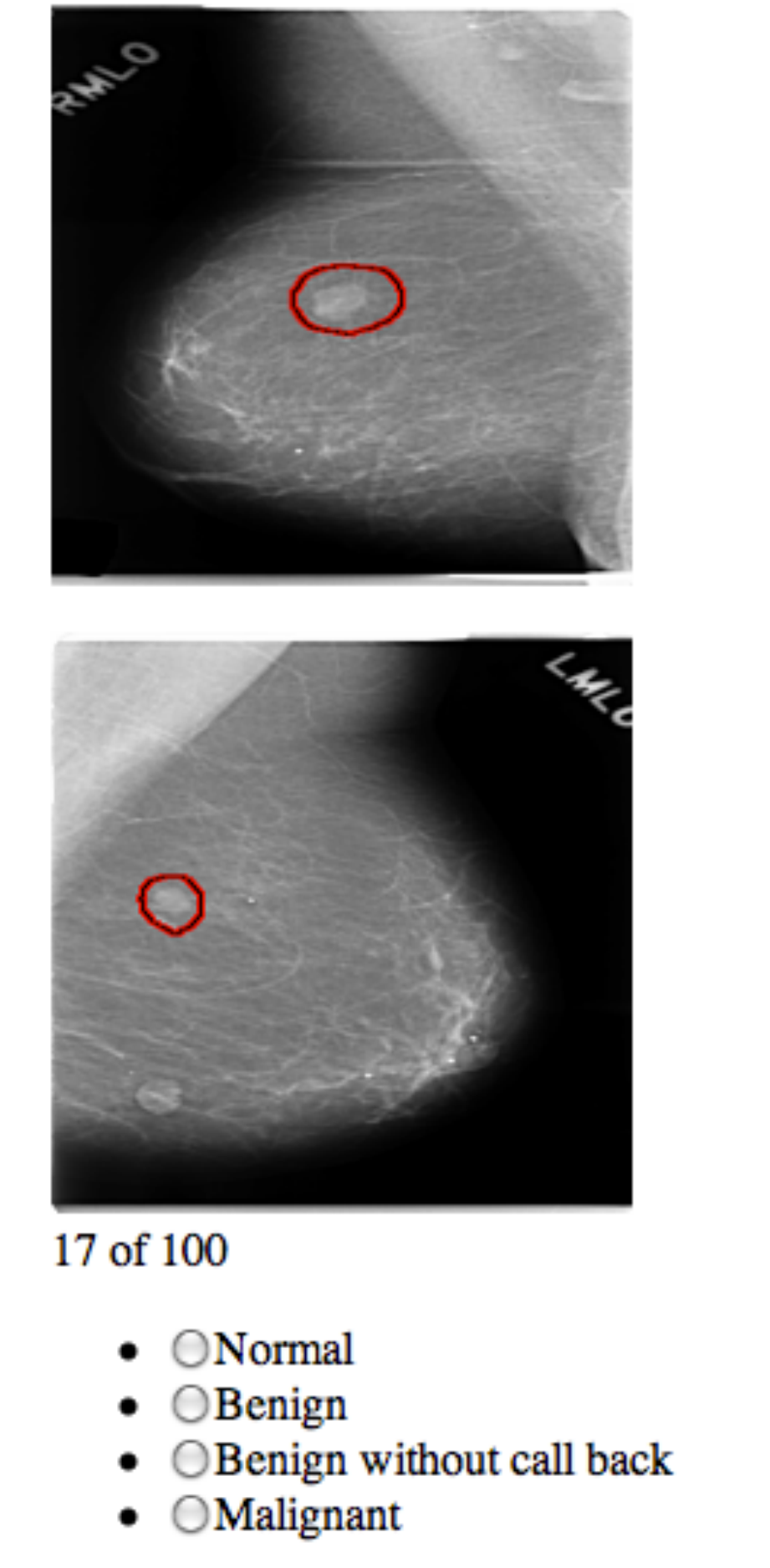}
\end{center}
\caption{\small {Example images of the Tumor Identification dataset. From left to right: First three are Malignant and fourth is benign.}}
\end{figure}

 \subsubsection{Results} This task clearly requires  expertise and is very difficult for an untrained person. On the other hand, an expert person can do really well. 
In this case, we had two labelers with less than  33\% mistakes and four labelers with more than 55\% mistakes. Results are shown in Table \ref{table3} and demonstrate once again the superiority of ELICE as compared to the other methods when injecting only 8 expert labels in the labeling process.


\subsection{Race Recognition Dataset}
\label{race_recognition}

\vspace*{0.2cm}

\begin{figure}[h]
\begin{center}
\includegraphics[scale=0.27]{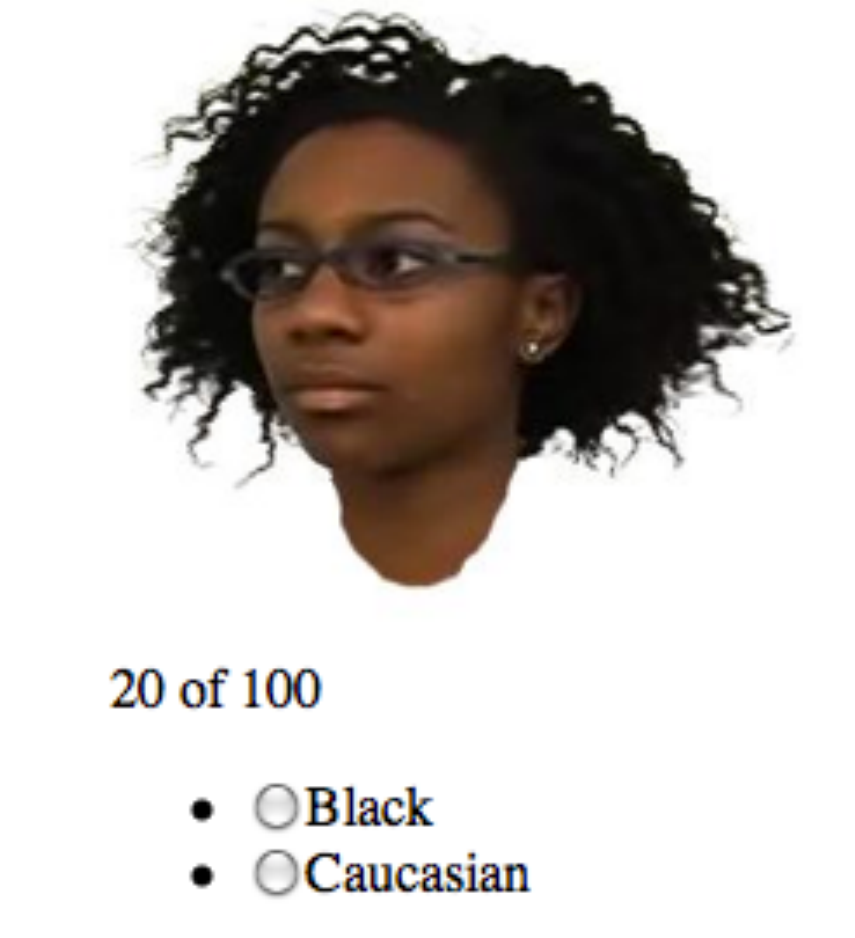}\   \includegraphics[scale=0.29]{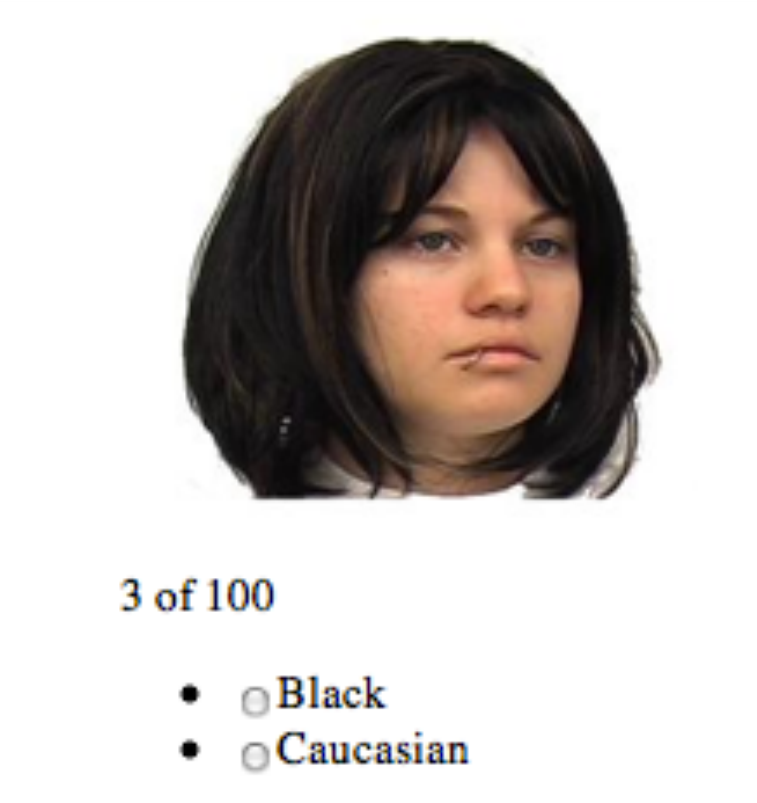}\   \includegraphics[scale=0.25]{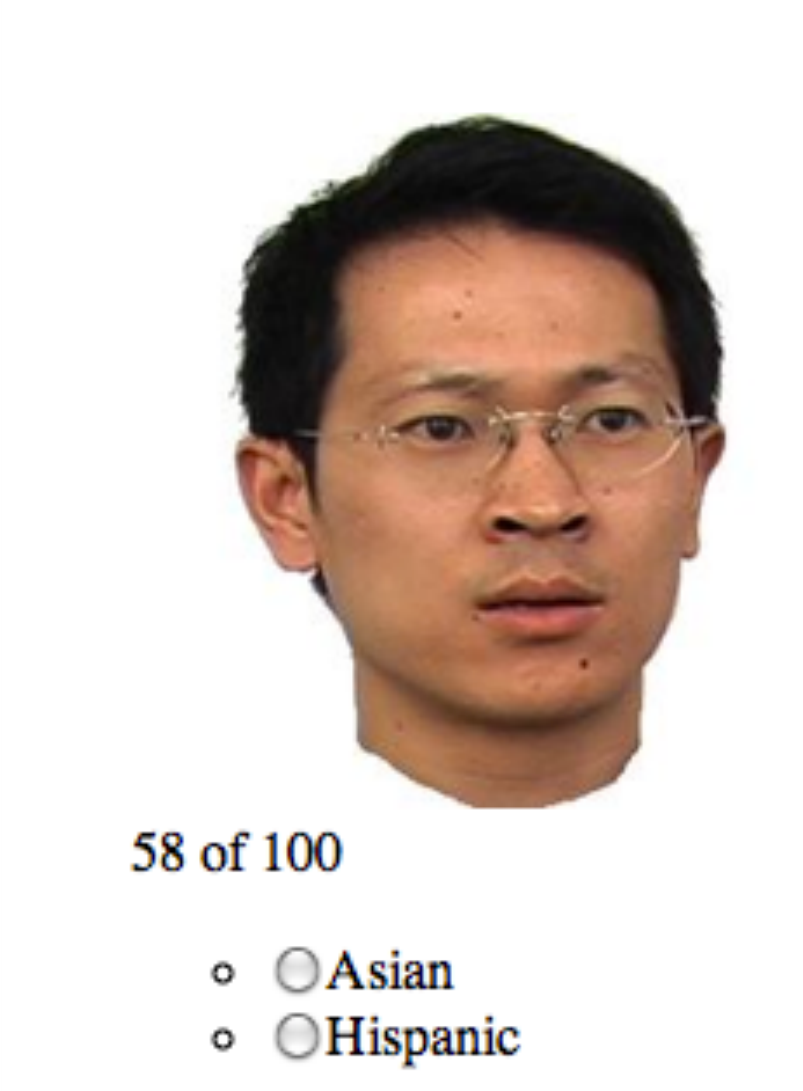}\     \includegraphics[scale=0.3]{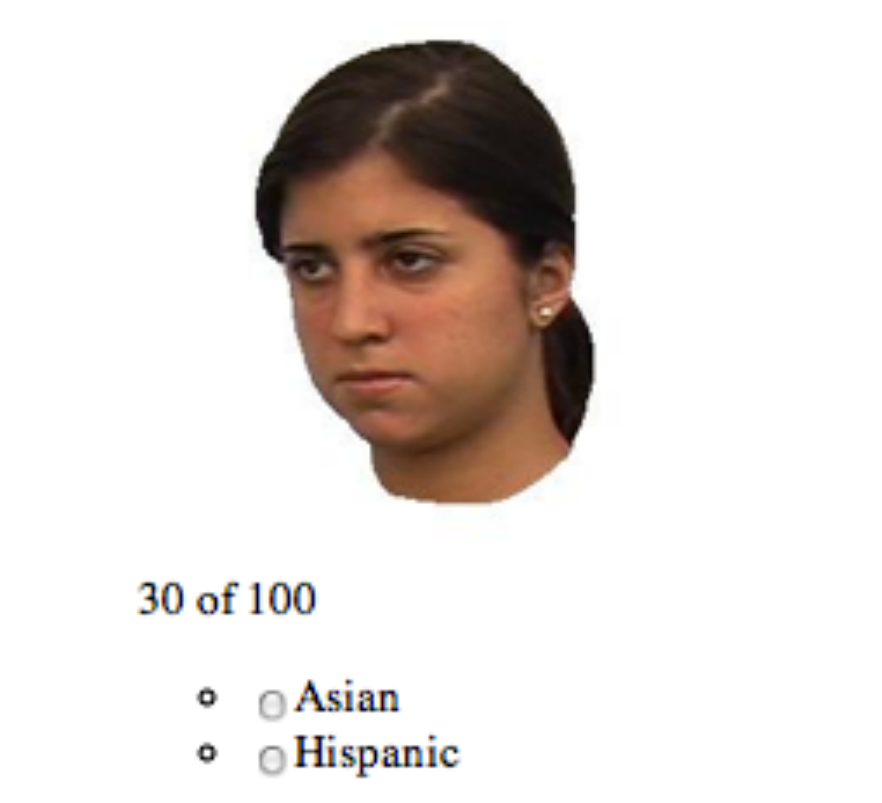}\\
\includegraphics[scale=0.29]{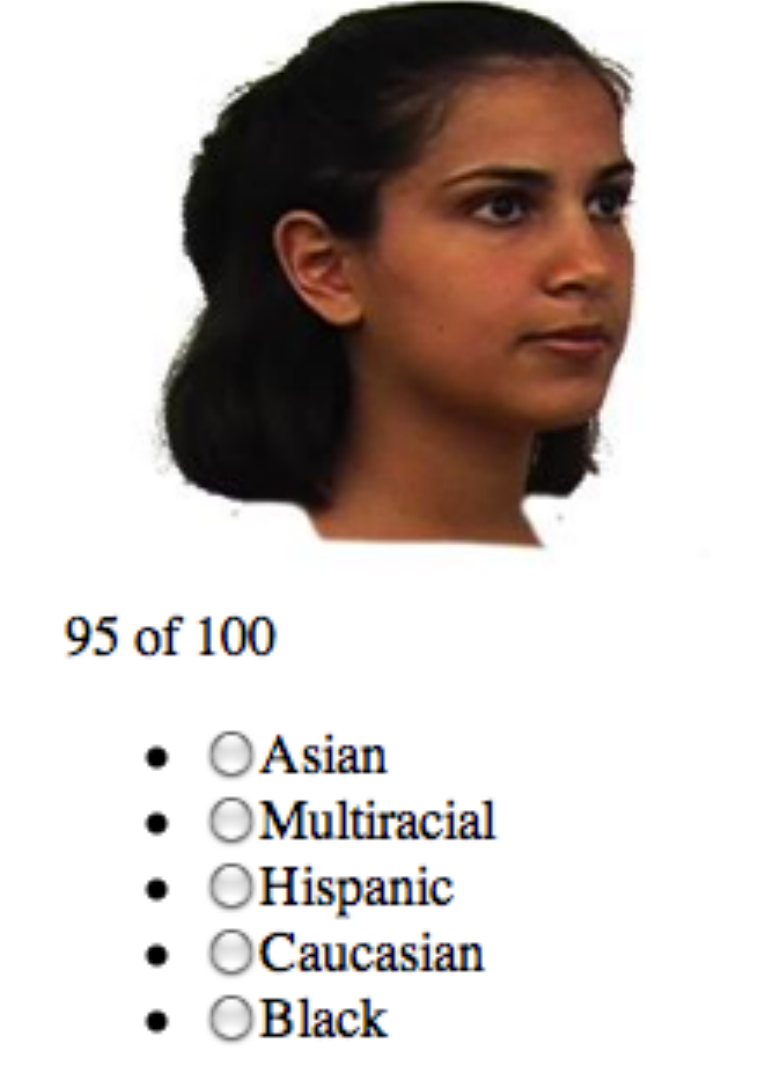}\   \includegraphics[scale=0.29]{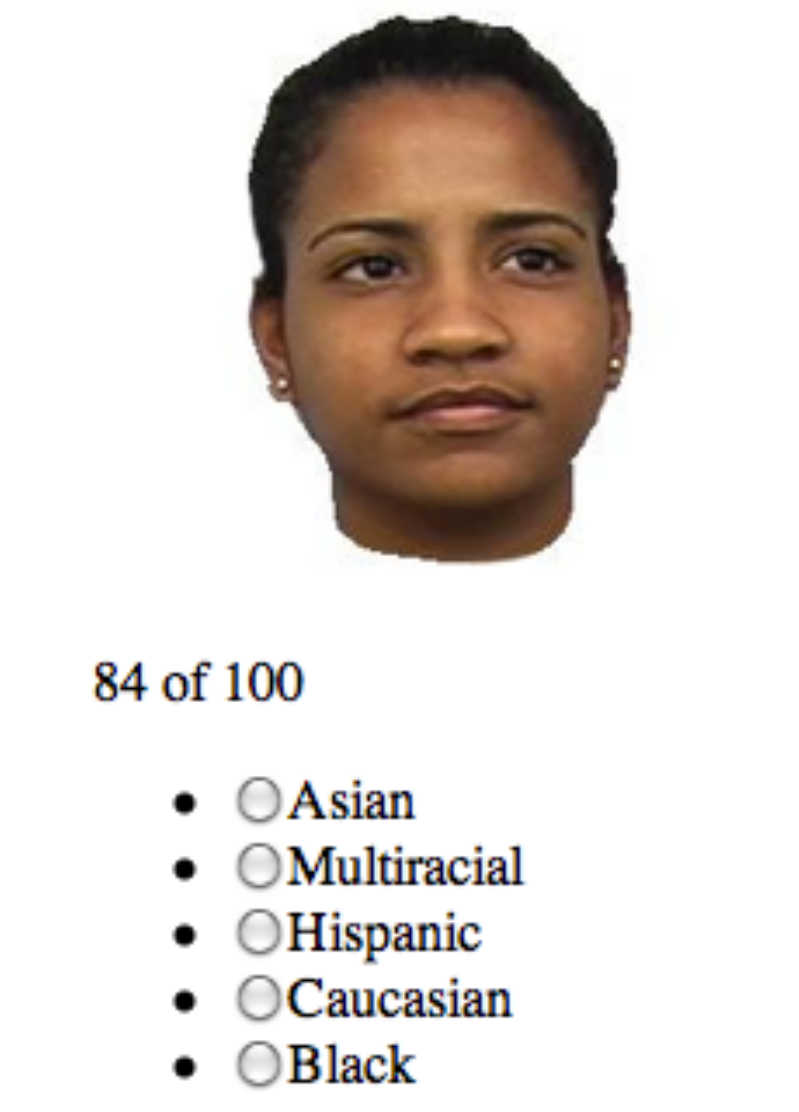}\    \includegraphics[scale=0.29]{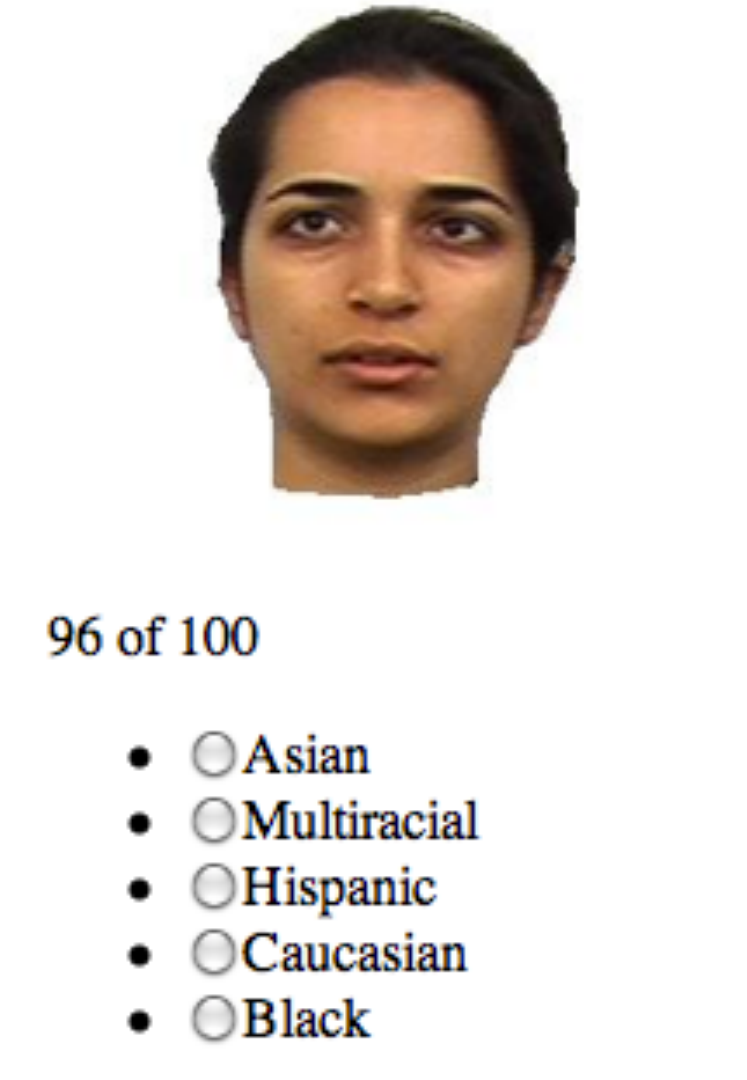}\    \includegraphics[scale=0.27]{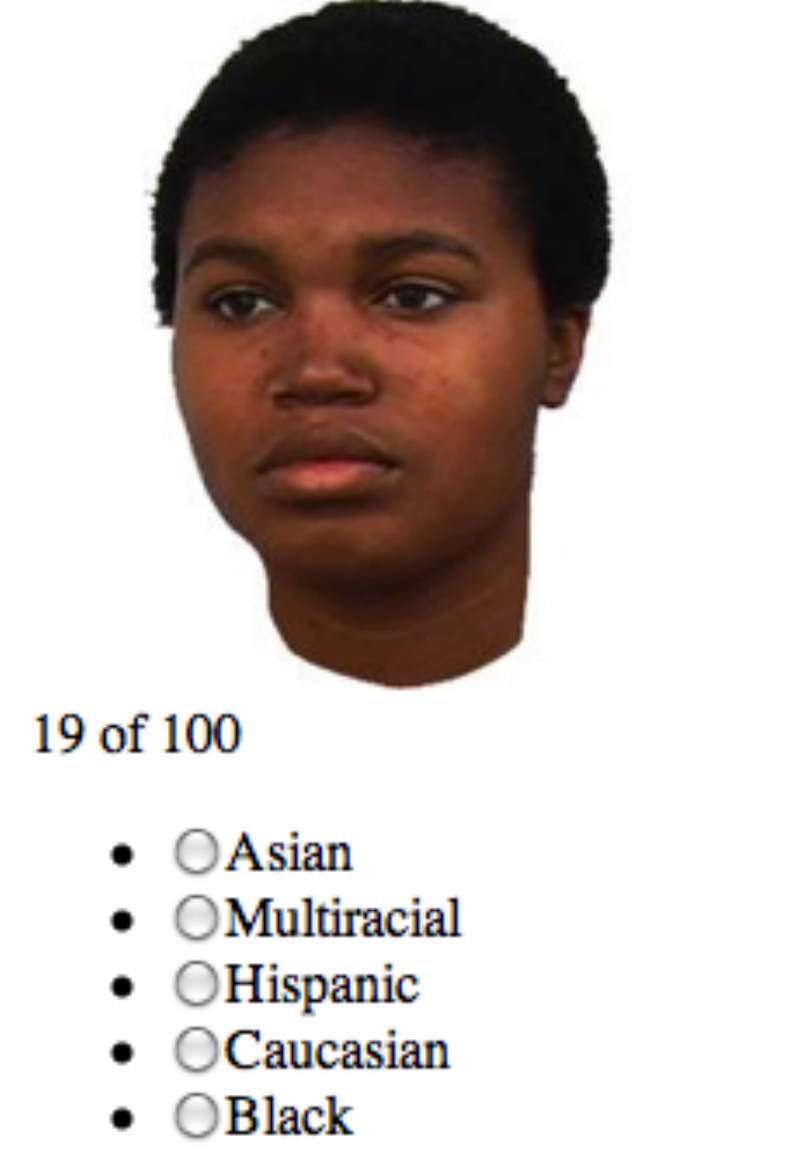}
\vspace*{-0.2cm}
\caption{\small{Example images from the Race recognition task posted on Amazon Mechanical Turk (Left to right): (Top) Black, Caucasian, Asian, Hispanic. (Bottom) Multiracial, Hispanic, Asian, Multiracial.}}
\label{race}
\vspace*{-0.5cm}
\end{center}
 \end{figure}

%

%
%

Another interesting real-life dataset we considered is {\em race recognition} dataset\footnote{Available on  Stimulus Images; Courtesy of Michael J. Tarr, Center for the Neural Basis of Cognition, Carnegie Mellon University \url{http://tarrlab.cnbc.cmu.edu/face-place}.} containing the images of people from different races. 

 \subsubsection{Experimental Design} We took three samples of 100 instances each and  posted them as a race recognition task on Amazon Mechanical Turk. The samples were chosen to guarantee different levels of difficulty. The tasks were to identify:
(1) Black versus  Caucasian (50 instances of each class),
(2) Hispanic versus Asian (50 instances of each class), (3) Multiracial versus other races (40 instances of Multiracial and 60 instances of the other races i.e. Asian, Black, Caucasian and Hispanic.) Snapshots of the experiment as posted on MTurk are shown in Figure \ref{race}.

For each task, we acquired six crowd labels for all 100 instances. The three tasks were chosen to guarantee easy to moderate difficulty level. \\

 \subsubsection{Results} 

 For all variants of ELICE, we used 8 random instances as expert-labeled instances. The results are shown in Table 2.
 Black versus Caucasian was the easiest of the tasks. Therefore, most of the labelers performed really well with only 0\% to 25\% of mistakes. As all the labelers had a good performance therefore the accuracy of all the methods was approximately perfect including the most naive method majority voting.

Identifying Hispanic versus Asian was relatively more difficult. In this case, some labelers made less than 15\% mistakes and the rest made over 48\% mistakes. In this case ELICE 2 performed best because of its ability to flip the labels.

The most confusing and challenging of all race recognition tasks was identifying multiracial from the other races. While most of the labelers did equally bad, surprisingly it was not as bad as we expected as the percentage of mistakes ranged between 30\% and 50\%. In this case almost all the labelers were falling in the random labeler category probably due to guessing rather than intelligently thinking the answer. In this case ELICE 1 was the winner but many other methods had approximately close results. The reason is that the random labelers do not provide much information. 


\begin{table*}[htbp]
{

\newcommand{\rr}{\raggedright} 
\newcommand{\tn}{\tabularnewline}

{\footnotesize
 \begin{tabular}{ l | c c c |c }                                         
  &  & \bf{Race Recognition} &&\bf{Tumor Identification}\\
                                          \hline
                                              &Black/Caucasian                    &Hispanic/Asian	                       &Multiracial/other       &Malignant/Non-malignant   \\
    \hline                                                         
   
{\bf MajorityVoting}			&0.9900	&0.5200	&0.6500				&0.5500\\	
{\bf GLAD	}				&{\bf 1.0000}	&0.5000	&0.6630			          &0.3043\\	
{\bf GLAD with Clamping}	  	&{\bf 1.0000}	&0.5000	&0.6630			          &0.3152\\	
{\bf Dawid Skene }              	& 0.9900    &  0.4500     	&0.6100	&0.7000\\  
{\bf EM }           	& 0.9900	&  0.5000		&0.6500	&0.3700\\ 
{\bf Belief Propagation 1}   	& 0.9900	&  0.5000		&0.6500	&0.3600\\
{\bf Belief Propagation 2 }   	& 0.9900	& 0.4500		&0.5900	&0.0600\\
{\bf Mean Field 1}               	& 0.9900	& 0.5000		& 0.6500 	&0.3600\\       
{\bf Mean Field 2}               	& 0.9900	& 0.4500		& 0.6000	&0.7000\\       
{\bf Karger 1}           	          	& 0.9900	&  0.5000		& 0.6500	&0.3600\\          
{\bf Karger 2 }                   		& 0.9900	&  0.5000		& 0.6500	&0.3600\\

{\bf ELICE 1}					&{0.9906}		&{0.6793}		&{\bf 0.6650} 	&{0.7100}\\	
{\bf ELICE 1 with clustering*}		&{-}&{-}		&{-}		         &{-}\\	
{\bf ELICE 2}			                &{0.9896} 		&{\bf 0.7648}   &0.5746	         &{0.7698}\\	
{\bf ELICE 2 with clustering*}	         &{-}&{-}&{-}&{-}\\	
{\bf ELICE Pairwise}			        &{ 0.9896} & 0.6729	&0.5756				& 0.7648\\	
{\bf ELICE Circular}			       &{ 0.9896}	& 0.6887	                         &0.5657		        &{\bf 0.7722}\\	
																															 \hline

  \end{tabular} }
\caption{\small{Accuracy of different methods on Amazon Mechanical Turk datasets. The given results are the  average of 100 runs on 100 instances with  6 labels per instance. Randomly chosen 8 instances are used as expert labeled instances (the instances with ground truth.) \hspace{18cm}
{\bf*}Since the features for these datasets are not available therefore the results of ELICE with clustering could not be calculated.  }}
\label{table3}
}
\end{table*}

 \section{Comparison and Discussion} 
 \label{comparison}
\vspace*{0.3cm}
 The  experimental evaluation shows superiority of ELICE as compared to the other state-of-the-art methods. 
   In this section, we will compare different versions of ELICE and discuss their appropriateness based on different situations.
  \begin{list}{$\bullet$}{} 
\item ELICE 1 is simple and easy to implement. It is not only efficient but also provides better results than the more complicated state-of-the-art methods. The key factor in this version of ELICE is that it relies on the judgment of the good labelers minimizing the effect of the random or malicious labelers. This can especially be helpful when at least one good labeler is available. When most of the labelers are average it may not produce very high accuracy but can still perform as good as the other prevailing methods. The low computational cost and effectiveness of the approach as compared to state-of-the-art methods are the main advantages of this version. This method can be used when the labeling task is not very challenging and there is a high chance to get at least one good labeler.
\item As compared to ELICE 1, the second version of ELICE provides even better accuracy because other than benefiting from good labelers, it takes also advantage of the malicious labelers.  
This is done through a better aggregation of labels that leads to incorporating the information from the malicious labelers.
This version is helpful when there is a high chance of the task being misunderstood or difficult resulting into unintentional malicious behavior. It can also take advantage of intentionally malicious labeler getting as much information as possible. While it is likely that not many labelers are intentionally malicious, whenever there is one the information provided is not wasted. 
 \item The third version of ELICE is based on the idea of incorporating most of the available information by comparison of labeler to labeler and instance to instance when ground truth is not known for certain. This version has a higher computational cost than our previous approaches, especially in the case of large datasets and should be used only in the case when  expert labels are not gold standard.
 \end{list}
ELICE framework is a one-shot method and does not block the labelers. 
Instead of keeping track of the labeler's history, we can simply estimate the ability of the labeler for one labeling task and improve accuracy, utilizing even the information provided by the malicious labelers. Although we do not completely disagree with the effectiveness of blocking the labelers, we believe that this technique may not be always helpful, mainly due to the following reasons:

  \begin{list}{$\bullet$}{} 
\item  A labeler can always do well on the test and poorly afterwards.
\item  A labeler may have more than one account and have different strategies on each of them. 
\item It is also possible that one account be used by more than one labeler at different times resulting into different performance levels. 
 \end{list}


%

\section{Bound on the number of expert labels}
\label{theorem2}
\vspace*{0.2cm}

\begin{figure}[ht]
\vskip 0.2in
\begin{center}
\includegraphics[scale=0.6]{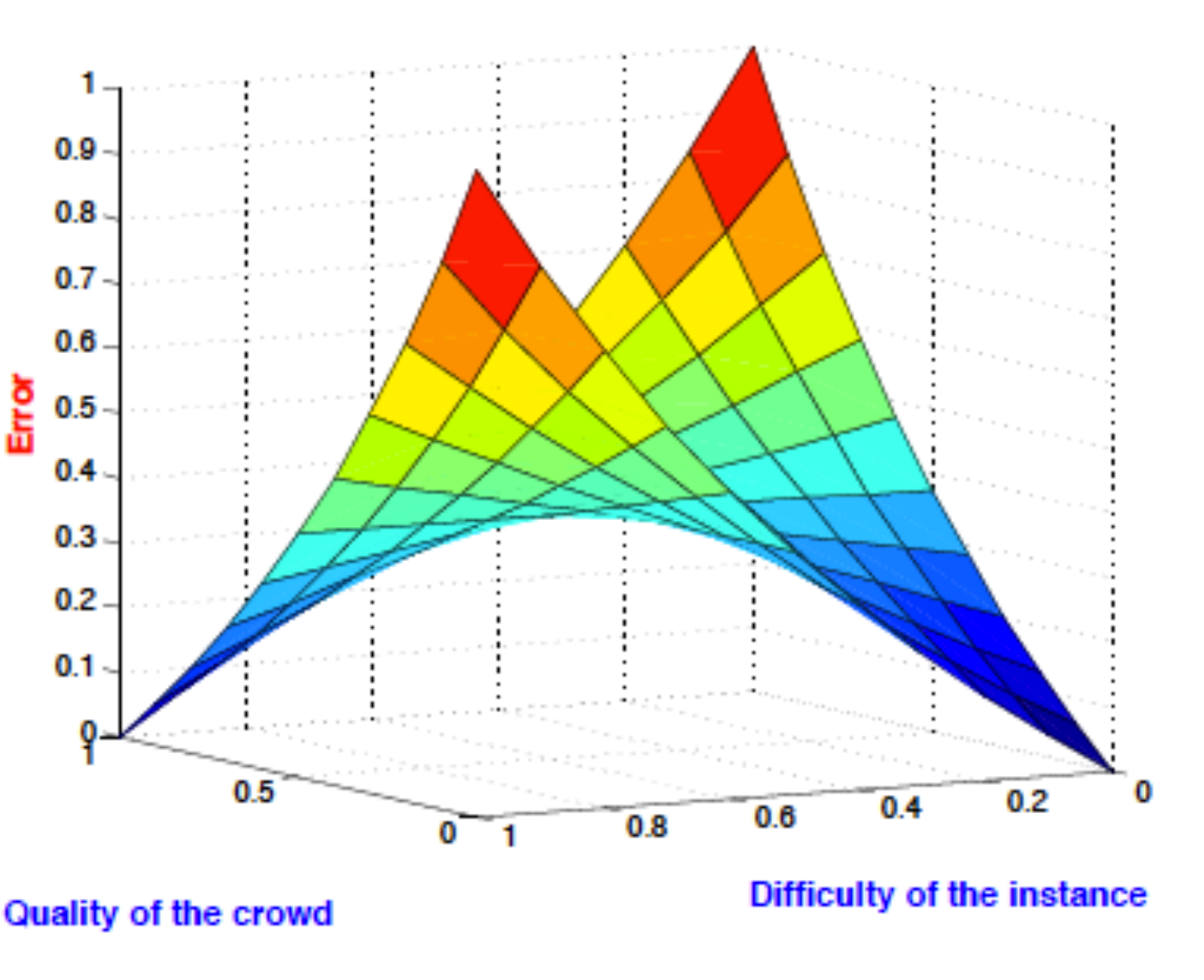}\    
\caption{Graph of the normalized error distribution. Quality of the crowd and difficulty of the dataset versus error.}
\label{errorfig} 
\end{center}
\end{figure}

\subsection{Motivation}
Expert-labeled instances are used to learn labeler expertise  $\alpha$  and difficulty of the instance $\beta$. Given that expert label acquisition can be expensive, it is desirable to find the lower bound of the expert-labeled instances needed, which can also provide a good estimate of $\alpha$ and $\beta$. This scenario is  similar to Probably Approximately Correct (PAC) learning where the learner has to learn the concept with minimum possible examples with a given accuracy and confidence. Therefore, we use the PAC learning framework to derive the bound.  As a prerequisite to this, we estimate the distribution of error in our  judgement of the crowd and instances. 

%
\subsection{Approach}
The error distribution depends on the overall quality of the crowd  and overall difficulty of the dataset, defined as follows:

 \begin{list}{$\bullet$}{} 
 \item{\bf Quality of the Crowd ($c$) : }
 Let $p_j $ be the probability of getting a correct label from labeler $j$ and $f$ be the probability distribution of $p_j$. Then we define the quality of the crowd $c$ as
$$c=E(P)= \sum_{j=1}^M \; p_j \; f(p_j)$$ 
Large values of $c$ represent better crowd.
 \item{\bf Difficulty of the Dataset ($1-d$) :}
We define, 
$$d=E(Q)= \sum_{i=1}^N \; q_i \; h(q_i)$$ 
where $q_i$ is the probability of getting the correct label for instance $i$ and $h$ is the probability distribution for $q_i$. Higher $d$ represents easier dataset.
\end{list}

 In general, $c$ and $d$ are unknown, we make a conjecture about the crowd quality and dataset difficulty based on the performance of crowd on a given dataset. So the error depends on how much the conjecture deviates from the true values of $c$ and $d$. Error is categorized as follows.
 \begin{list}{$\diamond$}{} 
 \item {\bf High}: When the crowd is good and we conjecture it as a bad crowd (or vice versa)  the error is high. This is also true when a dataset is easy and the conjecture is difficult (or vice versa).
   \item {\bf Medium}: When crowd is mediocre and we conjecture it as bad or good (or vice versa) then error is considered to be medium. Same is true about the dataset.
  \item {\bf Low}: Error is low when our judgment about the crowd and/or dataset is close to the true quality.
   \end{list}


\renewcommand{\arraystretch}{1.1}

  \begin{table*}[htbp]
{\small 
\begin{tabular}{|c|c|c |c| c| c| c|} 
\hline 

\multirow{2}{*}&& &\multicolumn{3}{|c|}{\bf Dataset} \cr 
\cline{4-6}
&&&{\bf Very Difficult }& {\bf  Moderate} & {\bf Very Easy} \\ [0.5ex] 

\hline 
\multirow{9}{*}{\begin{sideways}{\bf Crowd} \end{sideways} }
&\multirow{3}{*}{\bf Very Bad} &Conjecture about Crowd &Bad &Bad--Avg. &Avg.--Good \\ 
\cline{3-6}
&&  Conjecture about Dataset &Diff. & Diff.--Mod. &  Diff.-- Mod.-- Easy \\
 \cline{3-6}
 & &Error & Low & Medium & High \\
\cline{2-6}
\cline{2-6}
&\multirow{3}{*}{\bf Average} & Conjecture about  Crowd &Bad--Avg. & Bad--Avg.--Good & Good-- Avg. \\
\cline{3-6}
 && Conjecture about Dataset &Diff.--Mod. &Diff.--Mod.--Easy  & Mod.--Easy  \\
 \cline{3-6}
& & Error & Medium & Medium & Medium \\
 \cline{2-6}
\cline{2-6}
&\multirow{3}{*}{\bf Very Good}  &Conjecture about  Crowd & Bad-- Avg. & Good--Avg. & Good \\ 
\cline{3-6}
   &&Conjecture about Dataset & Diff.--Mod & Diff.--Mod. & Easy--Mod. \\
   \cline{3-6}
   & & Error & High & Medium & Low \\
 \cline{2-6}
\hline 
\end{tabular}
\vspace*{0.2cm}
\caption{Error distribution of the conjecture about the crowd and dataset. Crowd is categorized as very good, average, or very bad. Dataset is categorized as very easy, moderate, or very difficult. Error can be high, medium, or low.} 

\label{error_distribution} 
}
\end{table*}

The intuitive explanation of the error is summarized in Table \ref{error_distribution} and described as follows:

\begin{enumerate}[a)]
  \item {\bf Good crowd \& difficult instances}: When the crowd is good and instances are difficult the performance of the crowd may be average. The conjecture made is that the crowd is bad to average and/or the dataset is  of medium to high difficulty.  So the error is high in this case.
    \item {\bf Bad crowd \& difficult instances}: If  the crowd is very bad and instances are very difficult, then the  performance of the crowd will be poor. Hence the conjecture will be bad crowd and/or difficult dataset. Therefore, the error is low. 
    \item {\bf Good crowd \& easy instances}: When the crowd is very good and the instances are very easy our conjecture is good crowd and/or easy instances. Therefore, the error is low. 
   \item {\bf Bad crowd \& easy instances}: When the crowd is bad and dataset is very easy  then the judgement can be biased and the error can be high.
      \item {\bf Average crowd OR Moderate instances}: When the crowd is of average capability then for any kind of the instances the judgment may not be very far from the true value hence the error is medium. This also holds for average difficulty dataset and any kind of crowd. 
 \end{enumerate}

 We formalize the relationship between the crowd, the  dataset quality, and the error by the function:
  $$e=\frac{1}{1+(c-1/2)(d-1/2)}$$
 
  When the crowd is below average i.e. $c$ is less than $1/2$,  we have $(c-1/2)<0$. When crowd is above average  $(c-1/2)>0$. This is also true for $d$.
 When the values of c and d are close to 1/2  then $(c-1/2)(d-1/2)$ becomes small and hence $e$ becomes high. When the values of $c$ and $d$ are close to 0 or 1 $(c-1/2)(d-1/2)$ is relatively larger so $e$ is small. When one of the $c$ or $d$ is less than $1/2$ and the other is greater than $1/2$ then the value of $e$ is average. The graph of the function is shown in Figure \ref{errorfig}. 
 
%
%

 
%
%
%
%
%
%
%
%

\subsection{Theoretical Bound}  
For a given confidence $(1-\delta)$ and given values of $c$ and $d$, the lower bound on the number of expert labels is given by
 $$n\geq \frac{(b-a)(1+(c-1/2)(d-1/2))}{[1-a(1+(c-1/2)(d-1/2))]}log\; \frac{1}{\delta}$$
where $a$ and $b$ are the minimum and maximum of the values of the error $e$ respectively.

{\bf Proof:}
 The proof of this theorem is  straight forward. We know that the number of examples required by a PAC learning model is given by
  $$n \geq \frac{1}{\epsilon} log\; \frac{1}{\delta}$$
where $\epsilon$ is the error and $\delta$ is the level of confidence. In our case the error is depending on $c$ and $d$ hence the error $e$ is here 
  $$e=\frac{1}{1+(c-1/2)(d-1/2)}$$
  We normalize this error as follows
$$\epsilon=\frac{(e-a)}{(b-a)}$$
 where     $a$ = min$(e)$    \&        $b$ = max$(e)$ for $0<c<1$ and $0<d<1.$\\
 
 Therefore, we get 
 $$\epsilon={\frac{1}{(b-a)}[\frac{1}{1+(c-1/2)(d-1/2)}-a}]$$
Plugging the values in the PAC learning model we find the expression for the bound
$$n\geq \frac{(b-a)(1+(c-1/2)(d-1/2))}{[1-a(1+(c-1/2)(d-1/2))]}log\; \frac{1}{\delta}$$

\qed

\section{Related work}
\label{related_work}
\vspace*{0.2cm}

Many recent works have addressed the topic of {\em learning from crowd} (e.g., \cite{JMLR2010}).
The most common and straightforward approach to aggregate crowd labels is  majority voting. But the drawback of this method is that equal weights are assigned to the labels of each crowd labeler irrespective of his/her expertise. To overcome this problem,
 different solutions were proposed among which optimization methods using Expectation-Maximization (EM) are common.  In this context, Dawid \& Skene were the first to use the EM algorithm for finding better quality labels as well as approximating the expertise of the labeler \cite{Dawid:Skene:79}.
 
 A probabilistic model called Generative model of Labels, Abilities, and Difficulties (GLAD) is proposed by  \cite{WhitehillNIPS2009}. In their model, EM is used to obtain maximum likelihood estimates of the unobserved variables, which outperforms majority voting. The authors also propose a variation of GLAD that {\em clamps} some known labels into the EM algorithm. More precisely,  clamping is achieved by choosing the prior probability of the true labels very high for one class and very low for the other.
 
A probabilistic framework is also proposed by  \cite{YanAistat2010} as an approach to model annotator expertise and build classification models in a multiple label setting. 
A Bayesian framework is proposed by  \cite{RaykarICML2009} to estimate the ground truth and learn a classifier.  The main novelty of their work is the extension of the approach from binary to categorical and continuous labels. \cite{Sheng2008,Sorokin2008,Snow2008} show that using multiple, noisy labelers is as good as using fewer expert labelers. 

A second line of research (e.g., \cite{DonmezCS09}) uses active learning  to increase labeling accuracy by choosing the most informative labels. This is done by constructing a confidence interval called ``Interval Estimate Threshold'' for the reliability of each labeler.  Also, \cite{Yan_2011}  develop a probabilistic method based on the idea of active learning, to use the best  labels from the crowd. 

An iterative approach is proposed by  \cite{Karger,Karger3} which relies on a belief propagation algorithm to estimate the final labels weighted by each worker reliability.
Besides estimating the final labels, the method proposes an explicit approach of instance assignment to labelers using a bipartite graph generated by a random graph generation algorithm.
%
To handle adversarial labelers, \cite{Dekel2009} propose an approach that handles noisy labels. It is shown that the approach outperforms classification with Support Vector Machines,  especially when the noise in the labels is moderate to high.

Another approach aims to identify adversarial labelers (e.g., \cite{AMT}). This is tackled through an a priori identification of those labelers before the labeling task starts.  However, a malicious labeler can perform well initially and then adversarially behave during the labeling process. Similarly, the idea of using ground truth labels has been used in Crowdflower \cite{Edmonds} where  crowd ability is tested based on a few ground truth instances. 

The proposed approach tests the crowd labelers during the training phase (before the actual labeling starts) and blocks the labelers who do not pass the training.  Subsequent tests are also used to block bad crowd labelers after giving warnings. This is done by injecting instances for which ground truth is available during the actual labeling task. This approach can be helpful when a large number of ground truth instances are available. To handle this problem \cite{prog_gold} propose ``Programmatic gold"  that generates gold units automatically which may not be possible for many datasets.

The model  proposed by  \cite{Ipeirotis_qualitymanagement,IpeirotisP11} identifies  biased or adversarial labelers  and corrects their assigned labels. This is done by replacing \emph{hard labels} by a \emph {soft labels}.  Class priors and the probability of a labeler assigning an instance from a particular class to some other class is used to calculate the soft labels.

Our proposed method ELICE can handle adversarial and below average labelers, in an integrated way. Instead of identifying them separately, we propose to acquire few expert labels for instances that are representative of the data and judge the labelers  and instance difficulty {\em after} the labeling task is achieved. This method helps in getting good approximation even when good labels are not available either because of the difficulty of the task or because of inexperienced labelers.

\section{Conclusion}
\label{conclusion}
\vspace*{0.2cm}
With the advent of digitization, \emph{Big Data} became available everywhere, re-shaping almost every field in our daily life. While data is abundant, most of it still remain in an unlabeled form and not readily available for prediction tasks through machine learning algorithms. 
Crowd-labeling has emerged as an important line of research to help label sheer volumes of data to unlock its large potential. However, it remains crucial to harness the crowd efforts to get quality and reliable labels. 

In this paper, we have proposed a robust crowd-labeling framework using both expert evaluation and pairwise comparison between crowd-labelers. The framework embeds a set of methodologies to advance the state-of-the-art in crowd-labeling methods. Our methodologies are simple yet powerful and make use of a handful expert-labeled instances to squeeze the best out of the labeling efforts produced by a crowd of labelers.

 We propose a variety of methodologies to choose from according to the crowd characteristics and labeling needs. We show through several experiments on real and synthetic datasets that unlike other state-of-the-art methods, our methods are robust even in the presence of large number of bad labelers. One of the most important aspect of our method is overcoming the phase transition inherent in other approaches. We also derive a lower bound on the number of expert labels needed.

\vskip 0.2in
\bibliographystyle{plainnat}
\bibliography{arxiv}

\end{document}